\documentclass[preprint,11pt]{elsarticle}

\usepackage{amsmath}
\usepackage{amssymb}
\usepackage{amsthm}
\usepackage{pifont}

\usepackage{multirow}
\usepackage{array}
\usepackage{colortbl}
\usepackage{adjustbox}

\usepackage{graphicx}
\usepackage{float}
\usepackage{caption}
\usepackage{subcaption}
\usepackage{placeins}

\usepackage{enumitem}
\usepackage{multicol}
\usepackage[margin=3cm]{geometry}

\usepackage{verbatim}

\usepackage{hyperref}

\journal{Heliyon}
\begin{document}

\begin{frontmatter}
\title{Hemodynamic Simulation in the Aortic Arch Under Anemic Diabetic and Healthy Blood Flow Conditions Using Computational Fluid Dynamics}

\author[a]{Farzana Akter Tina}
\author[a,b]{Hashnayne Ahmed\corref{cor1}}
\author[a]{Hena Rani Biswas}

\affiliation[a]{organization={Department of Mathematics, Faculty of Science and Engineering, University of Barishal}, 
            city={Barishal}, 
            postcode={8200}, 
            country={Bangladesh}}
            
\affiliation[b]{organization={Department of Mechanical and Aerospace Engineering, University of Florida}, 
            city={Gainesville}, 
            postcode={32611}, 
            state={Florida}, 
            country={USA}}

\cortext[cor1]{Corresponding author}

\makeatletter
\renewcommand\@makefnmark{}
\makeatother

\fntext[]{Email(s): 
\href{mailto:farzanatina17@gmail.com}{farzanatina17@gmail.com} (F.A. Tina),
\href{mailto:hashnayneahmed17@gmail.com}{hashnayneahmed17@gmail.com} (H. Ahmed),  
\href{mailto:biswas.hena@yahoo.com}{biswas.hena@yahoo.com} (H.R. Biswas).}  

\fntext[]{ORCID(s): 
\href{https://orcid.org/0009-0008-2147-9942}{0009-0008-2147-9942} (F.A. Tina),
\href{https://orcid.org/0000-0002-2136-6816}{0000-0002-2136-6816} (H. Ahmed),
\href{https://orcid.org/0000-0002-8862-6248}{0000-0002-8862-6248} (H.R. Biswas).}

\begin{abstract}
This study investigates the hemodynamic behavior of blood flow in the aortic arch across anemic, diabetic, and healthy conditions using computational fluid dynamics (CFD) simulations with a non-Newtonian Carreau viscosity model. Velocity fields, pressure distributions, and wall shear stress (WSS) patterns were analyzed to assess the impact of blood rheology and vessel geometry. Anemic blood, with low viscosity and hematocrit, produced smooth, low-resistance flow with reduced WSS and pressure gradients, potentially impairing perfusion. Diabetic blood exhibited elevated viscosity, leading to increased flow resistance, higher WSS, and localized separation at arterial branches—conditions associated with vascular stress and remodeling. Healthy cases showed balanced hemodynamic behavior with localized flow acceleration but maintained physiological ranges. These findings highlight the mechanistic links between rheological properties and cardiovascular stress, supporting the role of CFD in non-invasive vascular risk assessment and motivating future integration of patient-specific data and structural modeling for enhanced clinical relevance.
\end{abstract}


\end{frontmatter}

\section{Introduction} \label{sec:introduction}

Understanding blood flow dynamics in the aortic arch is essential for advancing cardiovascular diagnostics, risk stratification, and therapeutic planning. As a critical conduit for oxygenated blood to the brain and upper extremities, the aortic arch exhibits complex hemodynamic behavior driven by its curved geometry, branching vessels, and pulsatile flow. These features generate swirling streamlines, recirculation zones, and spatial gradients in wall shear stress —all of which are mechanistically linked to pathologies such as atherosclerosis, aneurysms, and hypertension~\cite{Ross1999, Malek1999, Numata2016}. Conventional imaging techniques often fall short in resolving these dynamic flow structures, whereas computational fluid dynamics offers a high-fidelity approach to simulate and analyze vascular hemodynamics in detail.

A growing body of literature has applied CFD to study aortic flow, demonstrating how vessel geometry and blood rheology shape hemodynamic forces. For instance, Numata et al.~\cite{Numata2016} showed that aortic dilation amplifies helical flow and increases the oscillatory shear index (OSI), revealing cerebroprotective effects from regional perfusion strategies. Prahl Wittberg et al.~\cite{PrahlWittberg2015} highlighted the influence of vessel anomalies and red blood cell (RBC) dynamics on WSS distribution. Miyazaki et al.~\cite{Miyazaki2017} validated turbulence models against 4D Flow MRI, while Markl et al.~\cite{Markl2016} emphasized the clinical importance of patient-specific CFD for visualizing flow asymmetries and identifying recirculation zones. Ahmed et al.~\cite{Ahmed2024} explored the effect of disease conditions—such as anemia and diabetes—on hemodynamic parameters, showing how pathological rheology can perturb flow uniformity and increase mechanical stress.

Building on these efforts, subsequent studies have incorporated multiphysics effects and non-Newtonian rheology to improve physiological accuracy. Caballero and Laín~\cite{Caballero2014} compared Newtonian and Carreau–Yasuda models, concluding that shear-thinning behavior is critical for resolving near-wall flow. Silva et al.~\cite{Silva2024} investigated fluid–structure interaction (FSI) in post-endograft flows, showing how device implantation alters WSS and introduces turbulence. Additional studies across carotid and coronary arteries have associated low WSS and oscillatory flow with plaque formation and arterial remodeling~\cite{Carvalho2021, Malek1999}. Yet, despite these advances, few studies have systematically compared multiple blood conditions—such as anemic, diabetic, and healthy states—within a unified CFD framework. This absence of comparative rheological insight in the aortic arch represents a key gap in current literature.

Recent investigations have also emphasized the role of geometric complexity, turbulence modeling, and experimental validation. Xu et al.~\cite{Xu2020} found that Large Eddy Simulation (LES) significantly improves WSS predictions in disturbed regions compared to laminar models. Zandvakili et al.~\cite{zandvakilizandvakili2024} and Liu et al.~\cite{Liu2023} illustrated how bifurcation geometry shapes vortical structures and local shear gradients. Experimental studies by Zimmermann et al.~\cite{Zimmermann2023} and Šeta et al.~\cite{eta2017} further confirmed the presence of complex secondary flows and pressure fluctuations in the arch. While these efforts have improved realism, they often focus on either anatomy or turbulence but not on the comparative influence of pathological blood rheology—especially in the context of pulsatile aortic flow.

In this study, we address this gap by conducting a rheology-aware CFD analysis of blood flow in the aortic arch, comparing anemic, diabetic, and healthy conditions within a consistent simulation framework. A Carreau viscosity model is used to represent shear-dependent behavior, and a pulsatile inlet condition is applied to capture physiologic flow dynamics. The simulation resolves transient velocity profiles, pressure distributions, and WSS patterns to assess how hematocrit and viscosity influence mechanical stress and flow uniformity. Unlike prior work, this study presents a direct comparison of pathological and physiological blood states under identical boundary and geometric conditions, allowing for clear interpretation of rheological effects on vascular mechanics.

The remainder of this article is organized as follows. \textsection~\ref{sec:physical_and_mathematical_framework} describes the physical and mathematical modeling framework, including governing equations, rheology, and boundary conditions. \textsection~\ref{sec:numerical_methodology} outlines the numerical approach, including geometry reconstruction, meshing, and solver configuration. \textsection~\ref{sec:results_and_discussion} presents the simulation results and interprets key trends in velocity, pressure, and WSS across blood conditions. Finally, \textsection~\ref{sec:conclusion} summarizes the major findings, limitations, and implications for future clinical and computational research.

\section{Physical and Mathematical Framework} \label{sec:physical_and_mathematical_framework}
The aortic arch is one of the most complex and clinically significant regions of the human arterial system, where the combination of curvature, bifurcations, and pulsatile blood flow gives rise to rich and varied hemodynamic behavior. Modeling such flow requires a geometry that faithfully reflects anatomical features, as well as a mathematical formulation capable of capturing unsteady, three-dimensional, and non-Newtonian effects. In this study, we focus on quantifying physiologically relevant flow structures—particularly wall shear stress—under pulsatile conditions within a realistic aortic arch geometry.

\begin{figure}[h!]
    \centering
    \begin{subfigure}{0.45\linewidth}
        \centering
        \includegraphics[height=7cm]{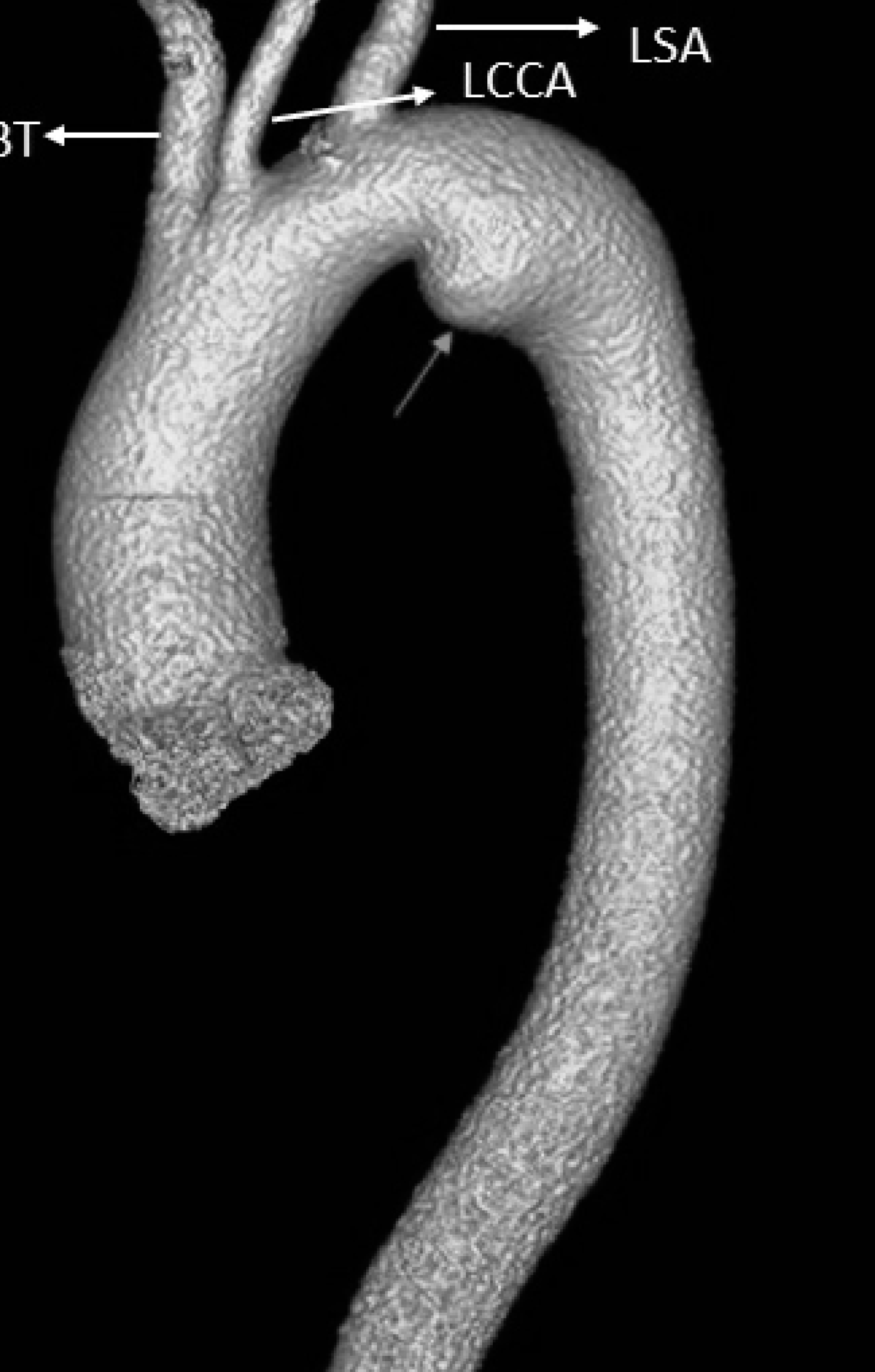}
        \caption{}
    \end{subfigure}
    \hfill
    \begin{subfigure}{0.45\linewidth}
        \centering
        \includegraphics[height=7cm]{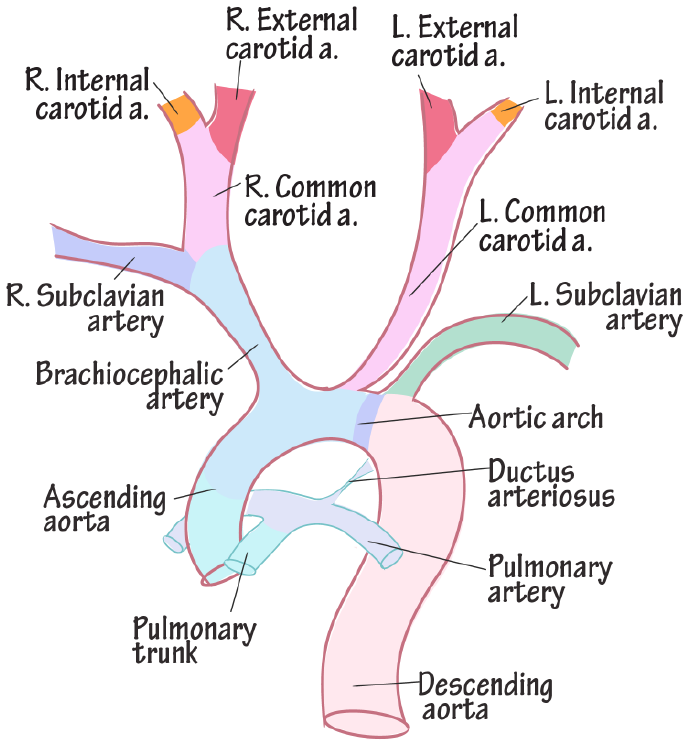}
        \caption{}
    \end{subfigure}
    \captionsetup{justification=justified, singlelinecheck=false}
    \caption{Visualization of the aortic arch. (a) MRI scan showing the anatomical structure of the aortic arch. Adapted from Baz et al. \cite{Baz2024}. (b) Schematic representation of the aortic arch highlighting key anatomical branches. Adapted from Natiq et al. \cite{Fadhil2023}.}
    \label{fig:arch_anatomy}
\end{figure}

To this end, the anatomical structure of the aortic arch is reconstructed from medical imaging data, capturing its major branches: the brachiocephalic, left common carotid, and left subclavian arteries. These branches are associated with strong geometric asymmetry and abrupt changes in curvature, both of which significantly influence the local distribution of shear forces and secondary flows. The geometry used in this study was obtained from a publicly available CAD model (\href{https://grabcad.com/library/human-aortic-arch-1}{https://grabcad.com/library/human-aortic-arch-1}) and refined to eliminate non-physiological surface irregularities. Figure~\ref{fig:arch_anatomy} shows the MRI scan and schematic diagram used to guide the modeling process. The geometry preserves critical branching angles and curvature radii that influence local hemodynamics.

From a physical modeling standpoint, the dynamics of blood flow in the aortic arch are governed by unsteady, incompressible fluid mechanics and by the non-Newtonian rheological behavior of blood. Because of the pulsatile nature of cardiac output and the geometric complexity of the domain, a time-dependent formulation of the Navier--Stokes equations is required. The governing equations for mass and momentum conservation are given as follows:
\begin{equation}
\nabla \cdot \mathbf{u} = 0
\label{eqn:mass}
\end{equation}
\begin{equation}
\rho \frac{\partial \mathbf{u}}{\partial t} + \rho (\mathbf{u} \cdot \nabla) \mathbf{u} = -\nabla p + \mu \nabla^2 \mathbf{u}
\label{eqn:momentum}
\end{equation}
where \( \mathbf{u} \) is the velocity field, \( p \) is pressure, \( \rho \) is the density of blood, and \( \mu \) is its effective dynamic viscosity. In physiological conditions, blood density typically ranges from \( 1050 \) to \( 1060 \, \text{kg/m}^3 \), while apparent viscosity varies from \( 0.003 \) to \( 0.16 \, \text{Pa}\cdot\text{s} \), depending on the local shear rate. These properties play a critical role in modulating both inertia and wall shear. The equations account for the inertial, pressure, and viscous forces acting on the fluid, with the unsteady term capturing the time-varying nature of cardiac output. To incorporate the shear-thinning behavior of blood, the dynamic viscosity \( \mu \) is defined using the Carreau--Yasuda model:
\begin{equation} \label{eqn:carreau}
\mu_{\text{eff}}(\dot{\gamma}) = \mu_\infty + (\mu_0 - \mu_\infty) \left( 1 + (\lambda \dot{\gamma})^2 \right)^{\frac{n-1}{2}}
\end{equation}
where \( \dot{\gamma} \) is the local shear rate, \( \mu_0 \) and \( \mu_\infty \) are the viscosities at zero and infinite shear rate, respectively, \( \lambda \) is a time constant, and \( n \) is the power-law index. This model effectively captures blood's shear-thinning behavior, particularly in the near-wall and low-flow regions common in curved arterial segments. Boundary conditions are prescribed to approximate in vivo hemodynamics. At the inlet of the ascending aorta, a pulsatile velocity waveform is applied to simulate periodic cardiac input. The waveform, based on the model by Owasit et al. \cite{Owasit2021, Ahmed2024}, is given by:
\begin{equation} \label{eqn:pulsatile_inlet}
v_{\text{inlet}}(t) =
\begin{cases} 
0.5 \sin \left[ 4\pi \left( t + 0.016 \right) \right], &  \text{if} \quad 0.5n < t \leq 0.5n + 0.218 \\
0.1, & \text{if} \quad 0.5n + 0.218 < t \leq 0.5(n+1)
\end{cases}
\end{equation}
where \( n = 0, 1, 2, \dots \), and the waveform spans a cardiac cycle duration of 0.5 seconds (corresponding to 120 beats per minute). This representation captures the acceleration and deceleration phases of systole and diastole. The Reynolds number based on peak systolic velocity and vessel diameter falls in the transitional regime (\( \sim 1000\text{--}2000 \)), indicating the potential for complex vortical structures. At the outlets of the supra-aortic branches, a constant static pressure of \( 100 \, \text{mmHg} \) (\( \sim 13332 \, \text{Pa} \)) is prescribed. Vessel walls are assumed to be rigid with a no-slip boundary condition \( \mathbf{u} = 0 \). While this approximation neglects wall compliance, it simplifies the problem and isolates the effect of flow pulsatility and geometric features. Future work may incorporate fluid–structure interaction (FSI) to assess the influence of wall motion.

To quantify near-wall hemodynamics, the wall shear stress,  \(\tau_{\text{wss}} \) is computed as the tangential force per unit area exerted by the fluid on the arterial wall, given by \(- \mu_{\text{eff}} (\partial v / \partial r)_{\text{wall}} \), where \( \mu_{\text{eff}} \) is the effective viscosity, \( v \) is the tangential velocity component, and \( r \) is the coordinate normal to the wall. For nondimensional interpretation, the skin friction coefficient, \(C_f \)  is calculated as \( \tau_{\text{wss}} / (0.5 \rho v^2) \), where \( \rho \) is the fluid density. To further assess the rotational complexity of blood flow in the aortic arch, we computed the helicity field, a scalar measure of the alignment between velocity and vorticity vectors~\cite{Katsoudas2025}. Helicity captures the presence of swirling and helical flow structures, which are prominent in curved and branching vessels and defined as,
\begin{equation}
\mathcal{H} = u \left( \frac{\partial w}{\partial y} - \frac{\partial v}{\partial z} \right) + 
              v \left( \frac{\partial u}{\partial z} - \frac{\partial w}{\partial x} \right) + 
              w \left( \frac{\partial v}{\partial x} - \frac{\partial u}{\partial y} \right),
\end{equation}
where \( u, v, w \) are the velocity components in the \( x, y, z \) directions, respectively. These quantities serve as critical indicators of endothelial shear exposure and are used throughout this study to identify regions of disturbed flow, shear imbalance, and vascular risk. Together, the physical and mathematical framework presented here enables a high-fidelity investigation of pulsatile, three-dimensional, non-Newtonian blood flow in the aortic arch. This formulation enables investigation of flow disturbances near branching points, which are often associated with early-stage atherosclerotic development.

\section{Numerical Methodology} \label{sec:numerical_methodology}
\begin{figure}[H]
    \centering
    \begin{subfigure}{0.45\linewidth}
        \centering
        \includegraphics[height=7cm]{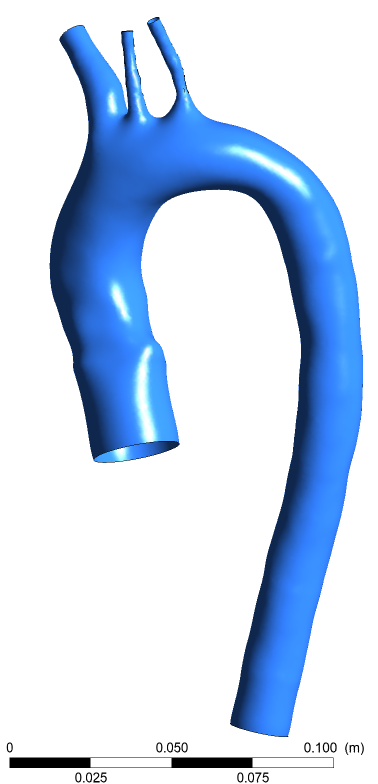}
        \caption{}
    \end{subfigure}
    \hfill
    \begin{subfigure}{0.45\linewidth}
        \centering
        \includegraphics[height=7cm]{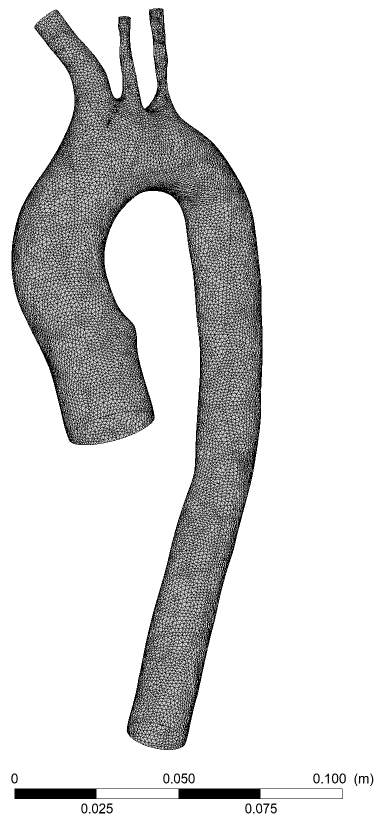}
        \caption{}
    \end{subfigure}
    \captionsetup{justification=justified, singlelinecheck=false}
    \caption{Reconstructed schematic of the aortic arch. (a) Three-dimensional model highlighting the main anatomical features of the aortic arch. (b) Mesh representation used in computational fluid dynamics simulations.}
    \label{fig:arch_reconstruction}
\end{figure}

To perform physiologically accurate simulations of blood flow in the aortic arch, the computational domain was constructed from a patient-inspired geometry, originally derived from medical imaging data and converted into a surface mesh in STL format. This file was preprocessed using CAD-based software to ensure anatomical consistency and compatibility with the meshing pipeline. Preprocessing operations such as smoothing and artifact removal were necessary to eliminate abrupt surface irregularities that could lead to numerical instability. This step ensures the final model remains both anatomically faithful and numerically robust. Figure~\ref{fig:arch_reconstruction} illustrates the reconstructed geometry and its discretized domain.

To verify that flow predictions were independent of spatial discretization, a grid sensitivity analysis was conducted. This test was crucial to ensure that numerical results—particularly velocity and wall shear stress (WSS)—remained consistent as the mesh was refined. Several meshes were tested by varying the maximum element size while holding all physical and boundary conditions constant. As shown in Table~\ref{table:grid_independence}, changes in key hemodynamic quantities diminished beyond a mesh size of \(1.0\, \text{mm}\), confirming convergence. The selected mesh achieves a balance between resolution and computational cost, enabling accurate yet efficient simulations. Hereafter, O1, O2, O3, and O4 refer to the brachiocephalic artery, left carotid artery, left subclavian artery, and descending aorta, respectively.
\begin{table}[H]
\centering
\caption{Grid independence study showing the average velocity at four outlets and wall shear stress (WSS) under increasing mesh resolutions. The bolded row corresponds to the mesh selected for all subsequent simulations.}
\label{table:grid_independence}
\resizebox{\textwidth}{!}{
\begin{tabular}{|c|c|cccc|cc|}
\hline
\multirow{2}{*}{\textbf{Elements}} & \multirow{2}{*}{\textbf{Max Size (mm)}} & \multicolumn{4}{c|}{\textbf{Average Velocity (m/s)}} & \multicolumn{2}{c|}{\textbf{WSS (Pa)}} \\
\cline{3-8}
 &  & \textbf{O1} & \textbf{O2} & \textbf{O3} & \textbf{O4} & \textbf{Average} & \textbf{Maximum} \\
\hline
48260   & 4.0   & 0.1680 & 0.1821 & 0.0953 & 0.0839 & 0.5941 & 4.7700 \\
72526   & 3.0   & 0.1708 & 0.1808 & 0.0916 & 0.0847 & 0.5731 & 4.6384 \\
98907   & 2.0   & 0.1539 & 0.1929 & 0.0909 & 0.0868 & 0.5382 & 3.6551 \\
\textbf{341047}  & \textbf{1.0} & \textbf{0.1547} & \textbf{0.2053} & \textbf{0.0989} & \textbf{0.0875} & \textbf{0.4950} & \textbf{3.9756} \\
417613  & 0.9   & 0.1523 & 0.2051 & 0.1046 & 0.0859 & 0.4870 & 3.6007 \\
527893  & 0.8   & 0.1502 & 0.1987 & 0.1044 & 0.0879 & 0.4841 & 2.9467 \\
690709  & 0.7   & 0.1543 & 0.1947 & 0.1072 & 0.0879 & 0.4763 & 3.0584 \\
\hline
\end{tabular}
}
\end{table}

Based on this evaluation, a final mesh was adopted comprising 341,047 unstructured tetrahedral elements and 502,590 nodes. Local refinement was applied in regions of high curvature and near the origins of the supra-aortic branches to resolve steep gradients in velocity and shear. Additionally, boundary layer elements were introduced near the vessel walls. This step is essential for accurately capturing wall shear stress, a key hemodynamic parameter analyzed in this study.

The simulations were performed using a transient incompressible flow solver, ANSYS Fluent 19.4R2 (\href{https://www.ansys.com/}{https://www.ansys.com/}). A pressure-based solver was selected, with the pressure–velocity coupling handled by the coupled scheme. This approach improves numerical stability, especially in complex geometries under unsteady conditions. Spatial derivatives were computed using the least-square cell-based method, and second-order upwind schemes were applied to the momentum equations. These higher-order methods reduce numerical diffusion and enhance accuracy in capturing convective transport. A hybrid initialization strategy was adopted to generate the initial flow field, combining patch and standard techniques to accelerate convergence. Residuals for continuity and momentum equations were required to drop below \(10^{-6}\) to ensure solution accuracy, with a maximum of 1000 iterations allowed per time step.

To validate the temporal resolution, a time-step sensitivity analysis was carried out. This analysis ensures that the temporal discretization does not distort the time-dependent solution. Six time steps were tested under constant total simulation duration, and their effects on peak velocity and WSS were evaluated. As shown in Table~\ref{table:time_step_sensitivity}, the variation across step sizes was negligible, confirming that a time step of \(0.4\, \text{s}\) is sufficient for resolving the pulsatile dynamics while keeping computational effort reasonable.

\begin{table}[H]
\centering
\caption{Time-step sensitivity analysis for a maximum mesh size of \(1.0\, \text{mm}\). The total simulation time was maintained at approximately \(10\, \text{s}\) across six different time step sizes. The bolded row corresponds to the selected time step used in all subsequent simulations.}
\label{table:time_step_sensitivity}
\resizebox{\textwidth}{!}{
\begin{tabular}{|c|c|cccc|cc|}
\hline
\multirow{2}{*}{\textbf{Time Step (s)}} & \multirow{2}{*}{\textbf{Total Steps}} & \multicolumn{4}{c|}{\textbf{Maximum Velocity (m/s)}} & \multicolumn{2}{c|}{\textbf{WSS (Pa)}} \\
\cline{3-8}
 & & \textbf{Outlet 1} & \textbf{Outlet 2} & \textbf{Outlet 3} & \textbf{Outlet 4} & \textbf{Average} & \textbf{Maximum} \\
\hline
0.6 & 17  & 0.2450 & 0.2944 & 0.2292 & 0.1753 & 0.4950 & 3.9756 \\
0.5 & 25  & 0.2450 & 0.2943 & 0.2292 & 0.1753 & 0.4950 & 3.9756 \\
\textbf{0.4} & \textbf{25} & \textbf{0.2450} & \textbf{0.2944} & \textbf{0.2292} & \textbf{0.1753} & \textbf{0.4950} & \textbf{3.9756} \\
0.3 & 33  & 0.2450 & 0.2943 & 0.2292 & 0.1753 & 0.4950 & 3.9756 \\
0.2 & 50  & 0.2450 & 0.2943 & 0.2292 & 0.1753 & 0.4950 & 3.9756 \\
0.1 & 100 & 0.2450 & 0.2943 & 0.2292 & 0.1753 & 0.4950 & 3.9756 \\
\hline
\end{tabular}
}
\end{table}

The Carreau--Yasuda model was employed to represent the non-Newtonian behavior of blood, capturing its shear-thinning response under varying flow conditions. A constant density of \(1060\, \text{kg/m}^3\) was used for all cases. Simulations were carried out for four distinct blood conditions—anemic, diabetic, and two healthy profiles—representing different rheological states. These conditions were selected to assess the effect of hematocrit and pathological viscosity changes on flow behavior. The specific parameters used in the model are shown in Table~\ref{table:blood_cases}.

\begin{table}[H]
    \centering
    \caption{Rheological parameters used in the Carreau--Yasuda model to represent anemic, diabetic, and healthy blood conditions, adapted from \cite{Attia2020, Ahmed2024}. These values capture the shear-thinning behavior of blood under different physiological states.}
    \label{table:blood_cases}
    \resizebox{0.85\textwidth}{!}{
    \begin{tabular}{|c|c|c|c|c|}
        \hline
        \multirow{2}{*}{\textbf{Properties}} & \textbf{Anemic} & \textbf{Diabetic} & \multicolumn{2}{c|}{\textbf{Healthy Blood}} \\ \cline{4-5}
        & \textbf{Blood} & \textbf{Blood} & \textbf{(Case 1)} & \textbf{(Case 2)} \\ \hline
        Power-law index, \( n \) & 0.33 & 0.39 & 0.48 & 0.3568 \\ \hline
        Constant time, \( \lambda \) (s) & 12.448 & 103.09 & 39.418 & 3.313 \\ \hline
        \( \mu_\infty \) (\( \text{Pa} \cdot \text{s} \)) & 0.00257 & 0.00802 & 0.00345 & 0.0035 \\ \hline
        \( \mu_0 \) (\( \text{Pa} \cdot \text{s} \)) & 0.0178 & 0.8592 & 0.0161 & 0.056 \\ \hline
        Hematocrit count & 25\% & 65\% & About 45\% & About 45\% \\ \hline
    \end{tabular}
    }
\end{table}

Each simulation was run for a total physical duration of \(10\, \text{s}\), covering multiple cardiac cycles and allowing full pulsatile behavior to be resolved. This setup, along with validated spatial and temporal resolution and physiologically accurate blood properties, provides a robust numerical framework for investigating hemodynamic phenomena in the aortic arch.

\section{Results and Discussions} \label{sec:results_and_discussion}
This section presents the simulation outcomes of blood flow in the aortic arch under four physiological and pathological conditions: anemic, diabetic, and two healthy cases. The results are analyzed through velocity distributions, pressure fields, and wall shear stress patterns to assess how blood rheology and geometry influence hemodynamic behavior. Each subsection focuses on a specific parameter, supported by both qualitative visualizations and quantitative comparisons across cases. These parameters are essential for understanding not only mechanical aspects of blood transport but also the onset of vascular dysfunctions such as atherosclerosis, hypertension, and thrombus formation. Integrating rheological properties with anatomical geometry helps bridge computational hemodynamics with clinical risk assessment~\cite{Steinman2003, cines1998endothelial}.

\subsection{Velocity Profiles and Flow Topology}

\begin{figure}[h!]
    \centering
    \begin{subfigure}{0.18\textwidth}
        \includegraphics[width=\linewidth]{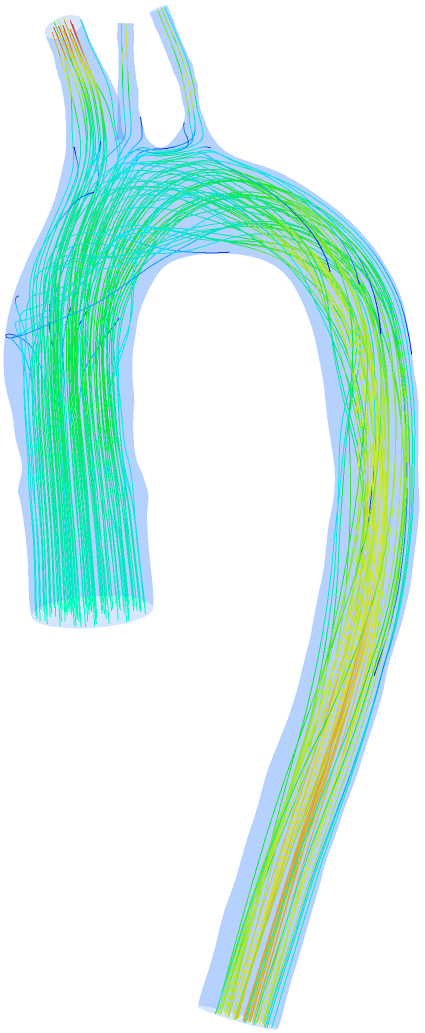}
        \caption{Anemic}
        \label{vs_a}
    \end{subfigure}
    \hfill
    \begin{subfigure}{0.18\textwidth}
        \includegraphics[width=\linewidth]{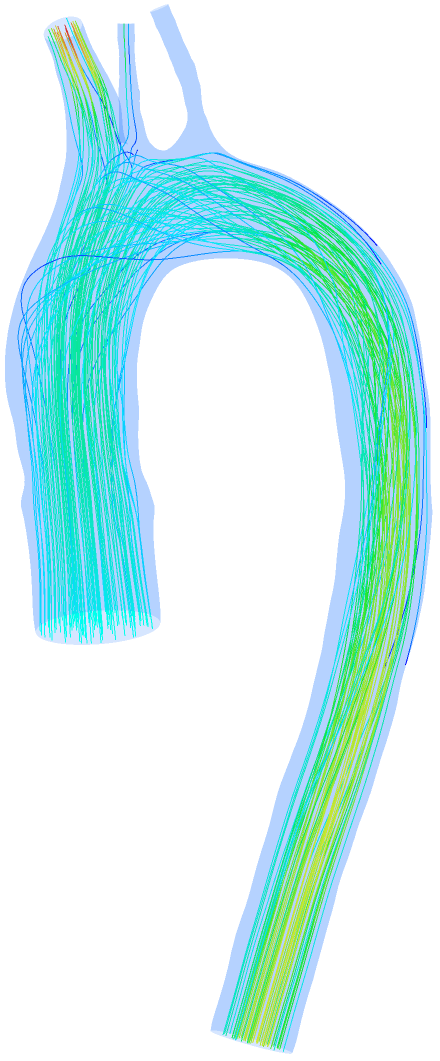}
        \caption{Diabetic}
        \label{vs_b}
    \end{subfigure}
    \hfill
    \begin{subfigure}{0.18\textwidth}
        \includegraphics[width=\linewidth]{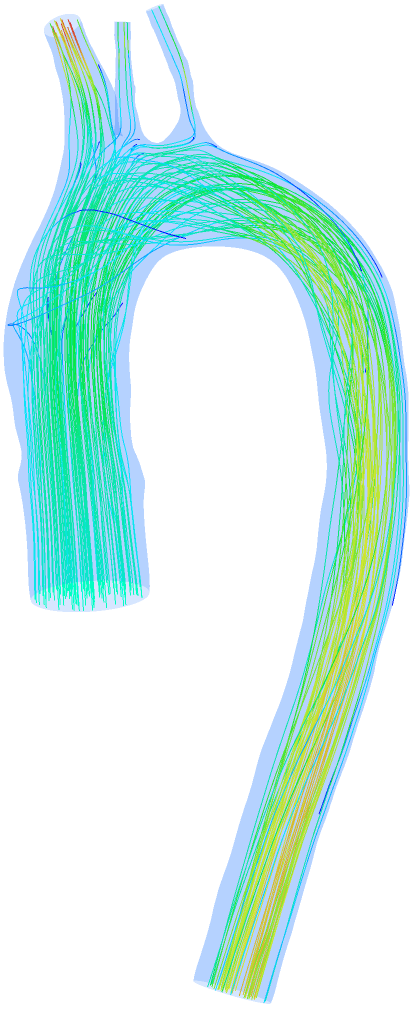}
        \caption{Healthy 1}
        \label{vs_c}
    \end{subfigure}
    \hfill
    \begin{subfigure}{0.18\textwidth}
        \includegraphics[width=\linewidth]{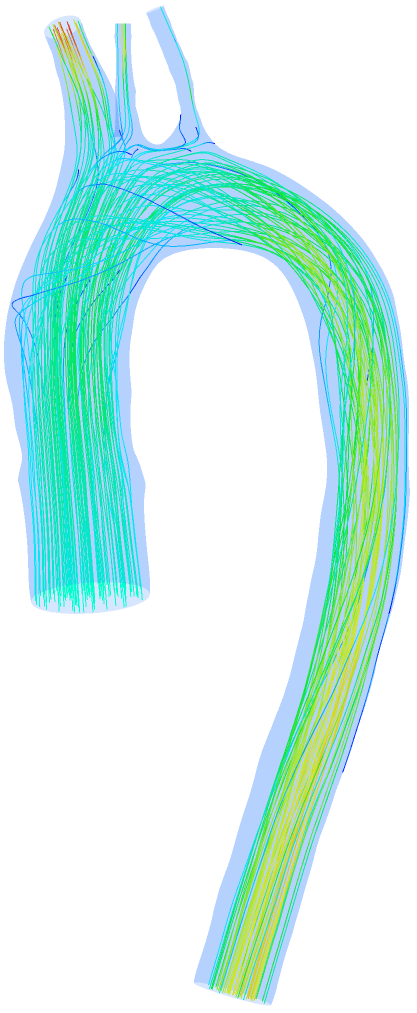}
        \caption{Healthy 2}
        \label{vs_d}
    \end{subfigure}
    \hfill
    \begin{subfigure}{0.08\textwidth}
        \includegraphics[width=\linewidth]{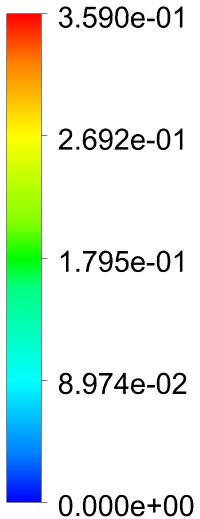}
        \caption*{}
    \end{subfigure}
    \captionsetup{justification=justified, singlelinecheck=false}
    \caption{Velocity streamlines in the aortic arch for different blood flow cases: (a) anemic, (b) diabetic, (c) healthy (Case 1), and (d) healthy (Case 2). Red regions represent higher velocity magnitudes; blue regions indicate lower values, in \( \text{ms}^{-1} \).}
    \label{fig:vs}
\end{figure}

\begin{figure}[h!]
    \centering
    \begin{subfigure}{0.18\textwidth}
        \includegraphics[width=\linewidth]{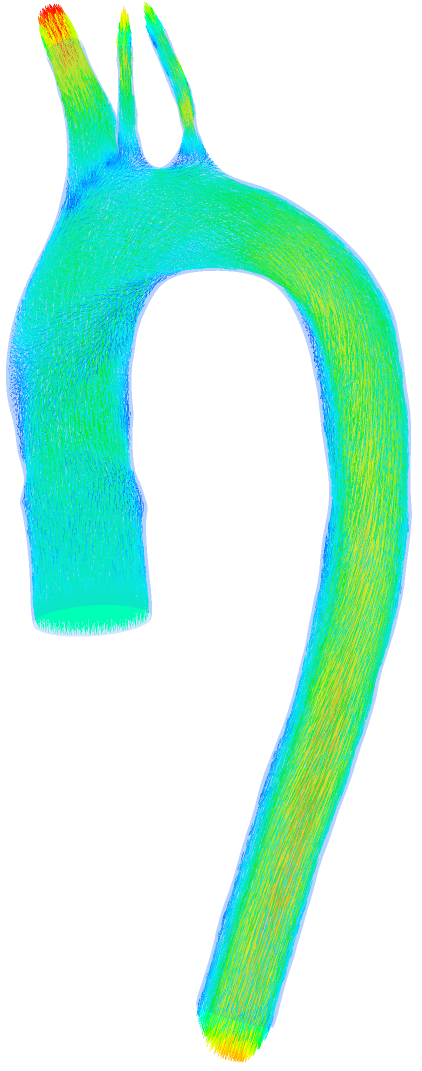}
        \caption{Anemic}
        \label{vv_a}
    \end{subfigure}
    \hfill
    \begin{subfigure}{0.18\textwidth}
        \includegraphics[width=\linewidth]{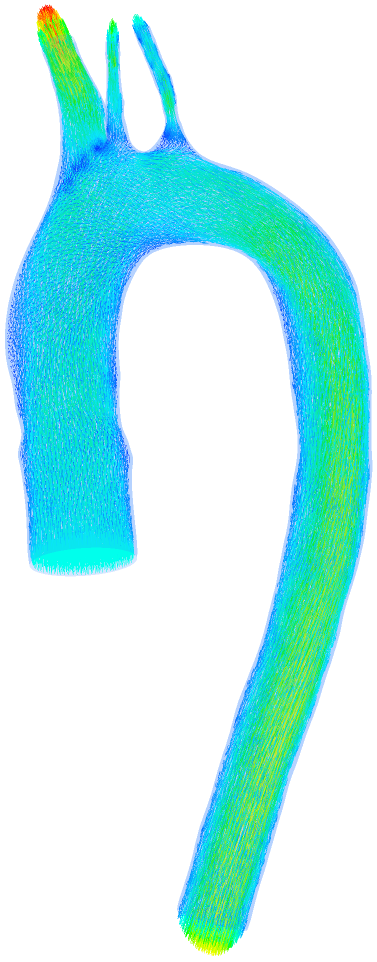}
        \caption{Diabetic}
        \label{vv_b}
    \end{subfigure}
    \hfill
    \begin{subfigure}{0.18\textwidth}
        \includegraphics[width=\linewidth]{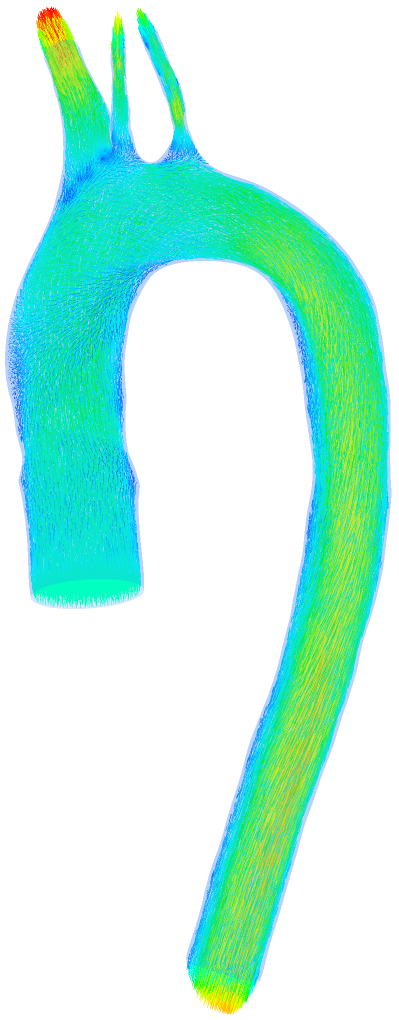}
        \caption{Healthy 1}
        \label{vv_c}
    \end{subfigure}
    \hfill
    \begin{subfigure}{0.18\textwidth}
        \includegraphics[width=\linewidth]{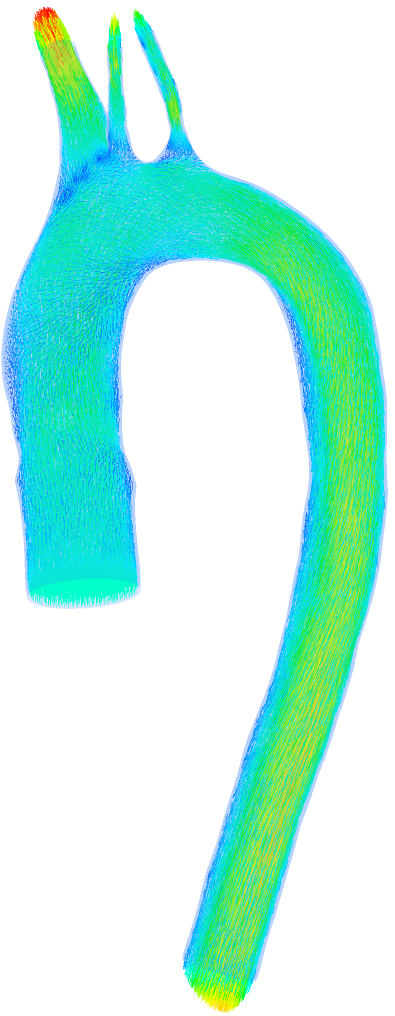}
        \caption{Healthy 2}
        \label{vv_d}
    \end{subfigure}
    \hfill
    \begin{subfigure}{0.08\textwidth}
        \includegraphics[width=\linewidth]{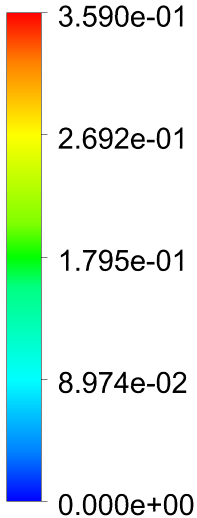}
        \caption*{}
    \end{subfigure}
    \captionsetup{justification=justified, singlelinecheck=false}
    \caption{Velocity vector fields in the aortic arch for different blood flow cases: (a) anemic, (b) diabetic, (c) healthy (Case 1), and (d) healthy (Case 2). Vector directions and magnitudes illustrate localized flow variations across conditions, in \( \text{ms}^{-1} \).}
    \label{fig:vv}
\end{figure}

Velocity distribution and flow topology are key determinants of hemodynamic performance in the aortic arch. Figures~\ref{fig:vs} and~\ref{fig:vv} compare velocity streamlines and vector fields, respectively, across four blood conditions: anemic, diabetic, and two healthy cases. These visualizations highlight how blood viscosity, hematocrit level, and vessel geometry shape overall flow characteristics. While streamlines emphasize bulk flow paths and recirculation, vector fields resolve local velocity magnitude and direction. Such flow patterns are physiologically significant because they modulate endothelial shear stress, particle residence time, and the likelihood of disturbed flow—all known to influence vascular health~\cite{cines1998endothelial, Morbiducci2010}. This is particularly relevant in curved arterial segments like the aortic arch, where centrifugal forces and secondary flow structures such as Dean vortices can amplify or suppress local disturbances, affecting both mechanical stress and particle transport~\cite{Morbiducci2011}.

In the anemic case (Figures~\ref{vs_a} and~\ref{vv_a}), the reduced viscosity and hematocrit yield a smooth, streamlined flow with well-aligned trajectories and minimal recirculation near branch points. The velocity vectors reinforce this, revealing uniform magnitudes and gradual directional changes. These features indicate low resistance and energy dissipation, reflecting the ease of blood transport in low-viscosity conditions. Clinically, this may reduce arterial wall stress but can impair effective perfusion in distal microcirculation if velocity becomes insufficient to sustain pressure gradients~\cite{Akinsheye2010}.

The diabetic case (Figures~\ref{vs_b} and~\ref{vv_b}) presents markedly more complex flow dynamics. Streamlines exhibit curvature and bifurcation-induced deflection, particularly along the inner aortic arch and near the brachiocephalic and carotid branches. Vector plots show steep gradients and localized high-velocity regions. These behaviors arise from elevated viscosity and hematocrit, which increase shear, reduce compliance, and introduce flow separation. The result is higher resistance and non-uniform flow behavior. Such disturbed flow topologies have been associated with endothelial dysfunction, altered nitric oxide production, and initiation of atherogenic processes in diabetics~\cite{Safar2017, Sun2014}.

Healthy Cases 1 and 2 (Figures~\ref{vs_c}, \ref{vs_d}, \ref{vv_c}, \ref{vv_d}) display intermediate patterns. Streamlines are generally smooth but reveal mild curvature at branch points due to natural geometric constraints. Velocity vectors remain balanced across the domain, indicating that blood transport is governed by geometry rather than pathological rheology. Case 1 shows slightly more flow acceleration near outlets, suggesting marginally higher viscosity or hematocrit compared to Case 2. These features are consistent with stable aortic flow that supports uniform wall shear distribution and minimizes mechanical stress on the endothelium.
\begin{figure}[h!]
    \centering
    \includegraphics[width=0.32\textwidth]{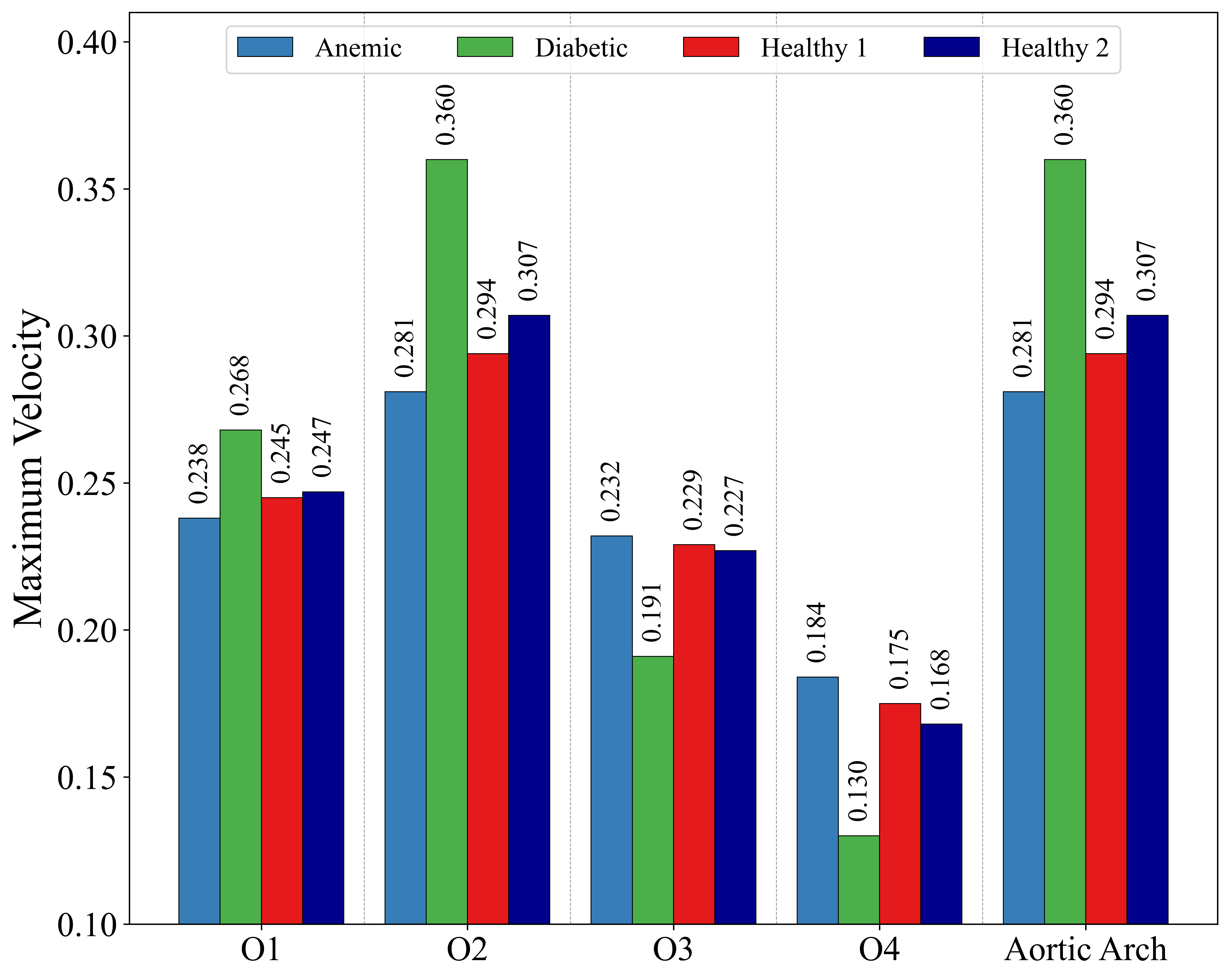}
    \includegraphics[width=0.32\textwidth]{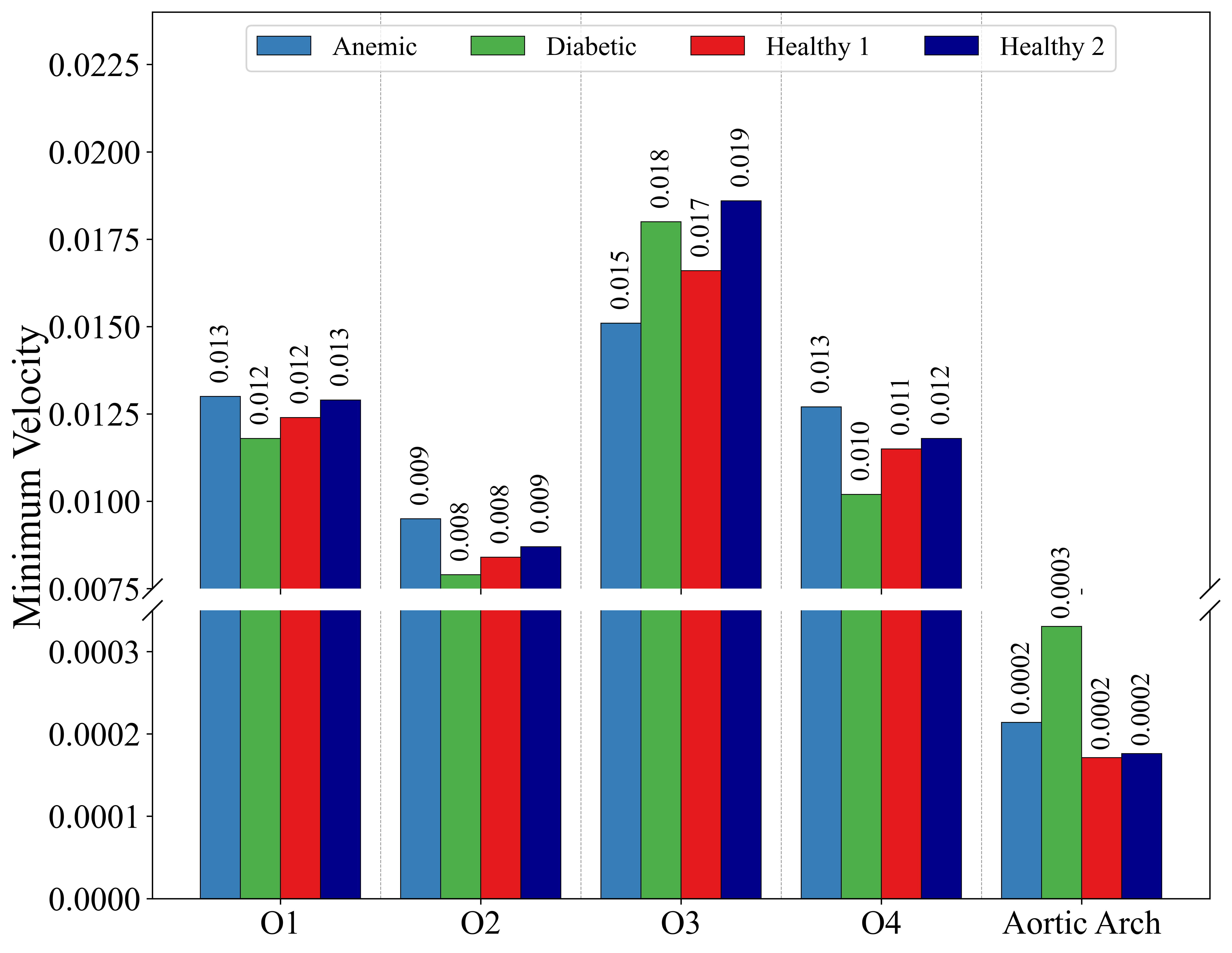}
    \includegraphics[width=0.32\textwidth]{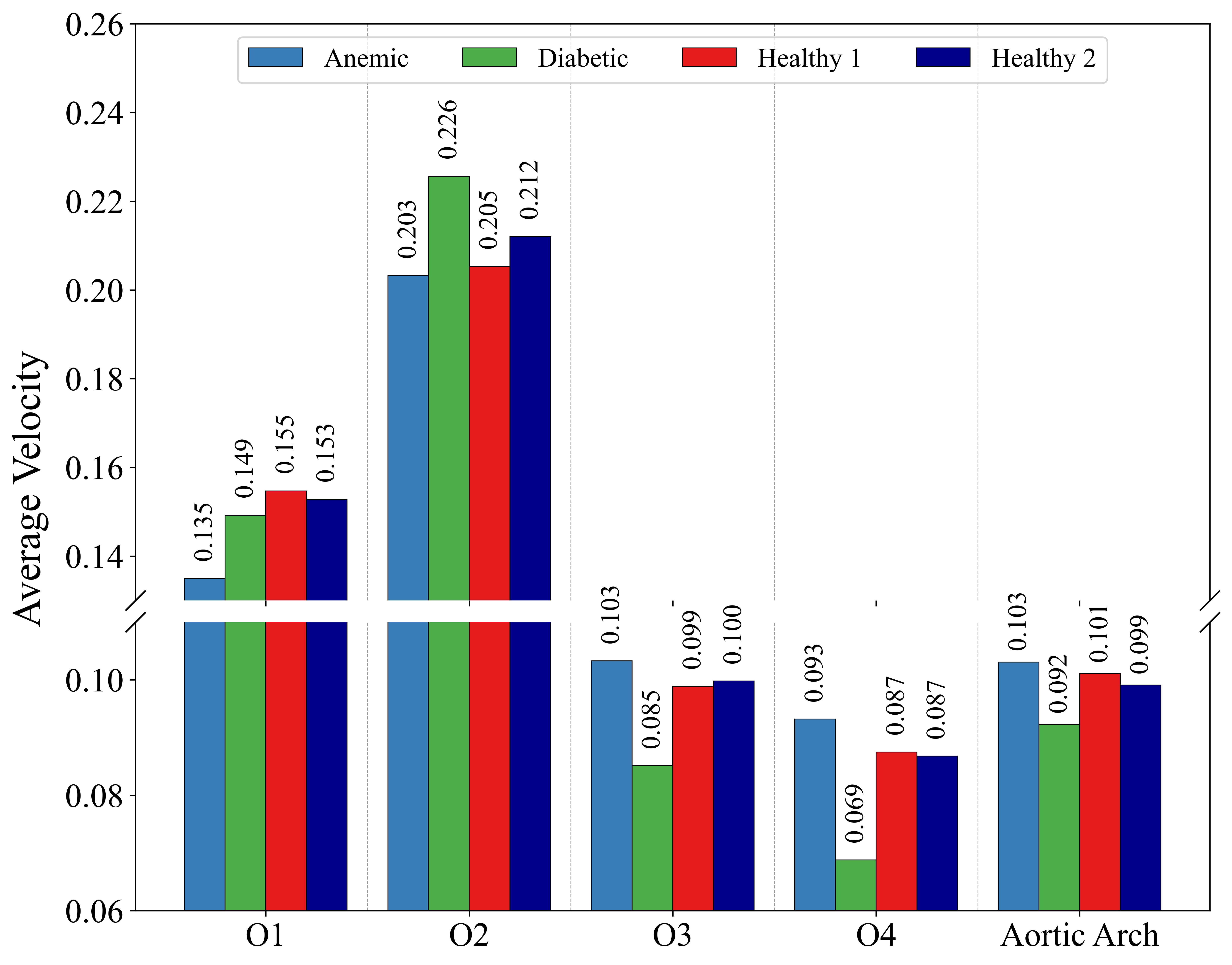}
    \caption{Velocity distributions: average (left), minimum (middle), and maximum (right) across different outlets and cases.}
    \label{fig:velocity-comparison}
\end{figure}

Figure~\ref{fig:velocity-comparison} quantifies the maximum, minimum, and average outlet velocities across all four blood conditions\footnote{O1, O2, O3, and O4 denote the brachiocephalic artery, left carotid artery, left subclavian artery, and descending aorta, respectively.}. Diabetic blood shows consistently elevated peak and mean velocities, particularly at O2, likely due to greater cardiac effort needed to overcome increased viscosity. Anemic flow exhibits the lowest velocities, reflecting reduced hematocrit and diminished transport capacity. Healthy cases cluster between these extremes, with Case 1 tending toward slightly higher values.

Minimum velocities show stronger contrasts. In diabetic and healthy flows, O3 and O4 demonstrate significant drops in minimum velocity, suggesting transient recirculation or localized flow deceleration—both clinically relevant for shear-induced platelet activation and potential thrombus formation~\cite{Steinman2003, Xu2020}. In contrast, the anemic profile remains relatively flat across all outlets, reflecting uniform forward flow with minimal resistance.

\begin{figure}[h!]
    \centering
    \begin{subfigure}{0.18\textwidth}
        \includegraphics[width=\linewidth]{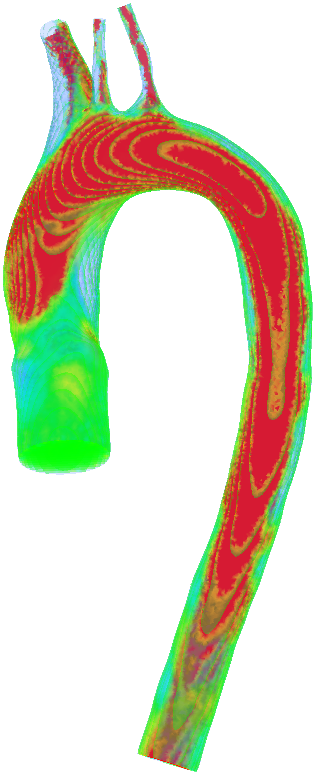}
        \caption{Anemic}
        \label{hl_a}
    \end{subfigure}
    \hfill
    \begin{subfigure}{0.18\textwidth}
        \includegraphics[width=\linewidth]{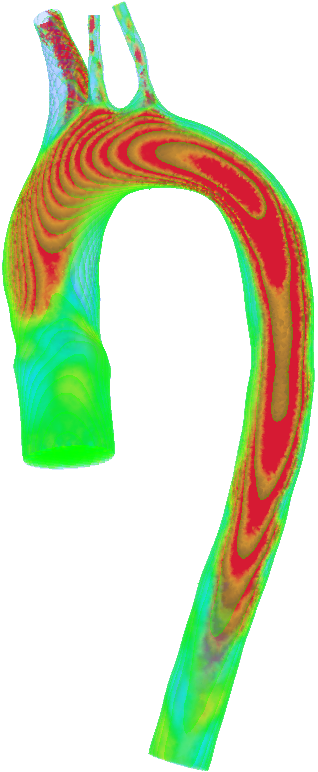}
        \caption{Diabetic}
        \label{hl_b}
    \end{subfigure}
    \hfill
    \begin{subfigure}{0.18\textwidth}
        \includegraphics[width=\linewidth]{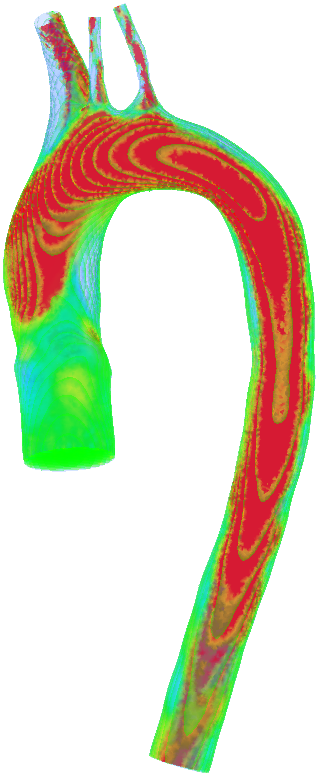}
        \caption{Healthy 1}
        \label{hl_c}
    \end{subfigure}
    \hfill
    \begin{subfigure}{0.18\textwidth}
        \includegraphics[width=\linewidth]{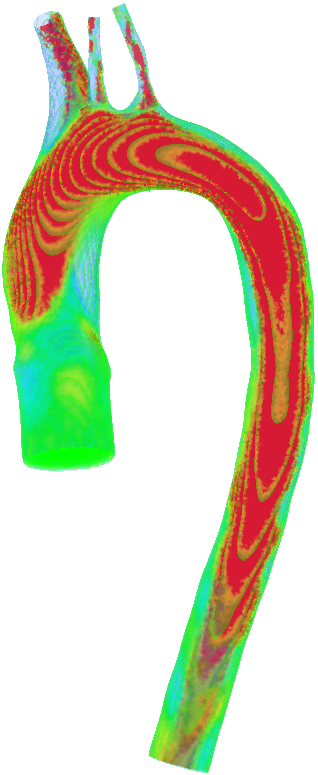}
        \caption{Healthy 2}
        \label{hl_d}
    \end{subfigure}
    \hfill
    \begin{subfigure}{0.08\textwidth}
        \includegraphics[width=\linewidth]{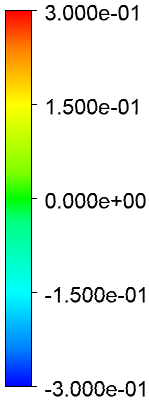}
        \caption*{}
    \end{subfigure}
    \caption{Helicity contours for four different patient-specific aortic flow cases: (a) anemic, (b) diabetic, (c) healthy (Case 1), and (d) healthy (Case 2). The color scale indicates helicity values in \( \text{m}^2\text{s}^{-2} \), with red and blue denoting strong positive and negative helicity regions, respectively.}
    \label{fig:hl}
\end{figure}

Figure~\ref{fig:hl} presents helicity distributions in the aortic arch for four physiological conditions: (a) anemic, (b) diabetic, (c) healthy (Case 1), and (d) healthy (Case 2). Helicity, defined as the scalar product of local velocity and vorticity vectors, quantifies the extent of rotational flow and is a marker of helical motion in pulsatile arterial environments. In the anemic case, helicity is substantially diminished, consistent with reduced blood viscosity and weakened secondary flow structures. The diabetic model shows enhanced helicity, especially near the arch curvature and supra-aortic branches, suggesting intensified vortical dynamics likely associated with increased flow resistance and vascular stiffening. Both healthy cases demonstrate moderate helicity concentrated along the outer wall, reflecting physiologically favorable swirling patterns that enhance wall shear stress distribution and promote efficient transport of oxygen and nutrients. These findings underscore the role of helicity as a biomechanical indicator of vascular health and pathology in cardiovascular simulations.

Taken together, these results show how pathological changes in blood properties significantly alter velocity distributions, with implications for nutrient delivery, shear stress exposure, and cardiovascular risk. The combined use of streamlines, vectors, and velocity statistics provides a robust framework for characterizing aortic hemodynamics across physiological states. This multidimensional approach also enables identification of vulnerable flow patterns that may predispose to endothelial dysfunction or vascular remodeling under disease conditions.

\subsection{Hemodynamic Pressure Distribution and Flow Resistance}

Pressure distribution in the aortic arch provides critical insights into the flow resistance and hemodynamic forces acting on the arterial walls.\footnote{The outlet pressure was set to 0\,Pa, and a time-varying velocity profile was imposed at the inlet. As a result, the simulation outputs gauge pressure (relative to the outlet). To obtain physiological (absolute) pressure values, a baseline offset of approximately 10{,}666\,Pa (i.e., 80\,mmHg) must be added.} Abnormal pressure gradients are clinically associated with increased afterload, impaired organ perfusion, and progressive cardiovascular disease, particularly in patients with hypertension or systemic vascular stiffening~\cite{Safar2017}.

\begin{figure}[h!]
    \centering
    \begin{subfigure}{0.18\textwidth}
        \includegraphics[width=\linewidth]{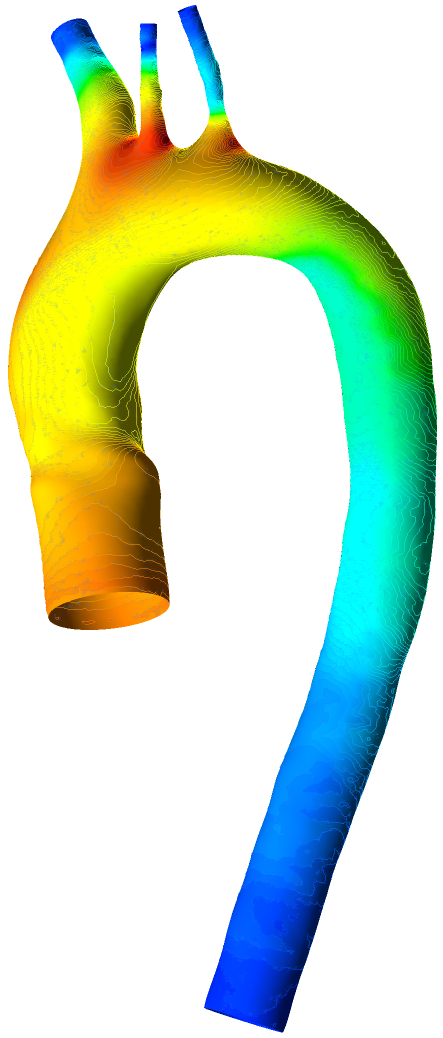}
        \caption{Anemic}
        \label{pc_a}
    \end{subfigure}
    \hfill
    \begin{subfigure}{0.18\textwidth}
        \includegraphics[width=\linewidth]{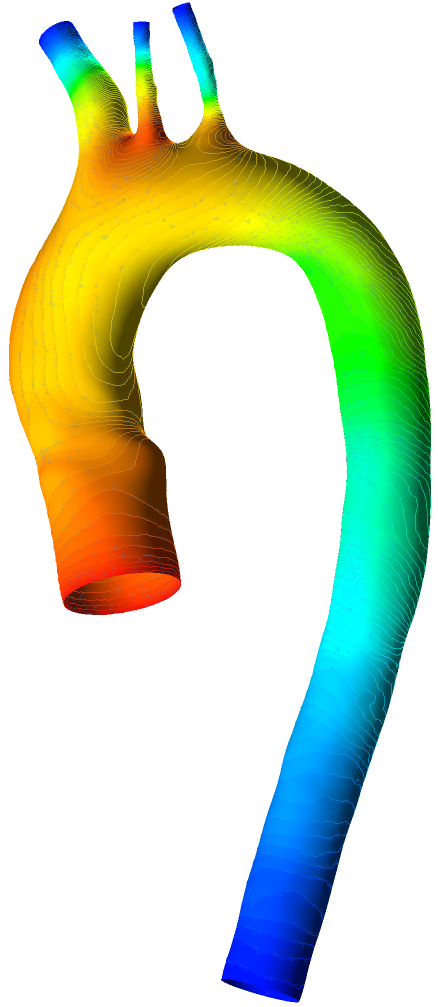}
        \caption{Diabetic}
        \label{pc_b}
    \end{subfigure}
    \hfill
    \begin{subfigure}{0.18\textwidth}
        \includegraphics[width=\linewidth]{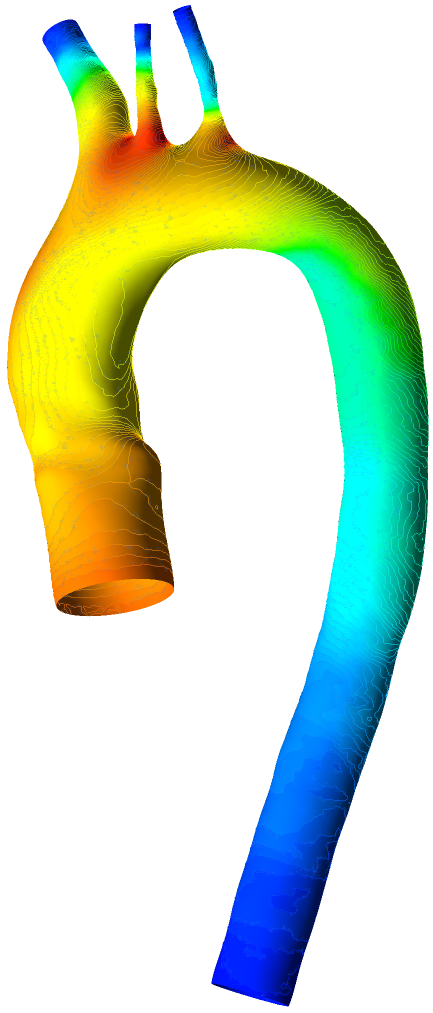}
        \caption{Healthy 1}
        \label{pc_c}
    \end{subfigure}
    \hfill
    \begin{subfigure}{0.18\textwidth}
        \includegraphics[width=\linewidth]{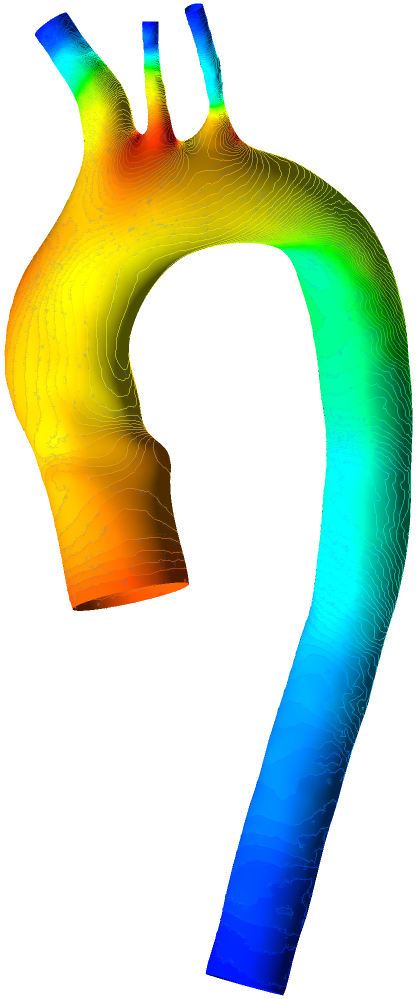}
        \caption{Healthy 2}
        \label{pc_d}
    \end{subfigure}
    \hfill
    \begin{subfigure}{0.08\textwidth}
        \includegraphics[width=\linewidth]{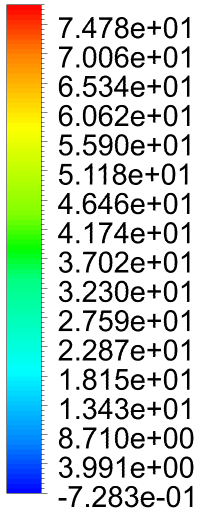}
        \caption*{}
    \end{subfigure}
    \captionsetup{justification=justified, singlelinecheck=false}
    \caption{Pressure contours in the aortic arch for different blood flow cases: (a) anemic, (b) diabetic, (c) healthy (Case 1), and (d) healthy (Case 2). Red regions correspond to areas of higher pressure, while blue regions indicate lower pressure, with values presented in Pascal (Pa), highlighting variations due to different blood conditions and physiological factors.}
    \label{fig:pc}
\end{figure}

\begin{figure}[h!]
    \centering
    \begin{subfigure}{0.18\textwidth}
        \includegraphics[width=\linewidth]{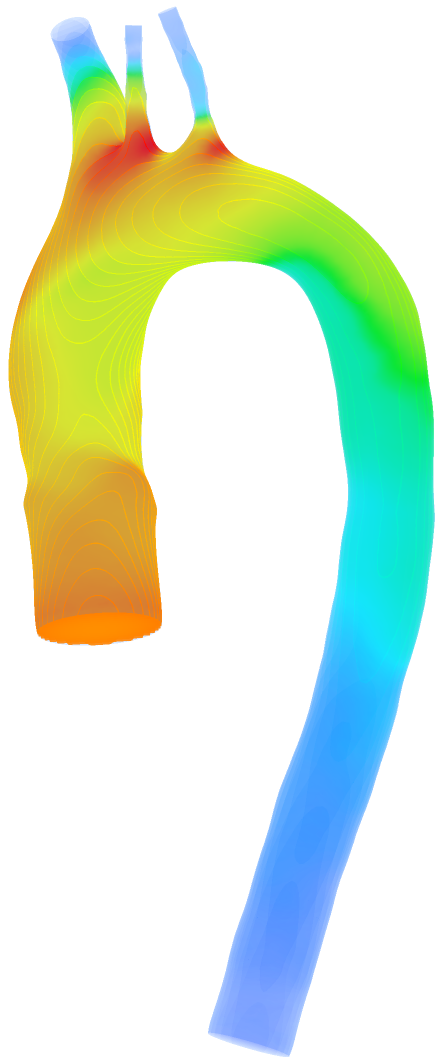}
        \caption{Anemic}
        \label{pr_a}
    \end{subfigure}
    \hfill
    \begin{subfigure}{0.18\textwidth}
        \includegraphics[width=\linewidth]{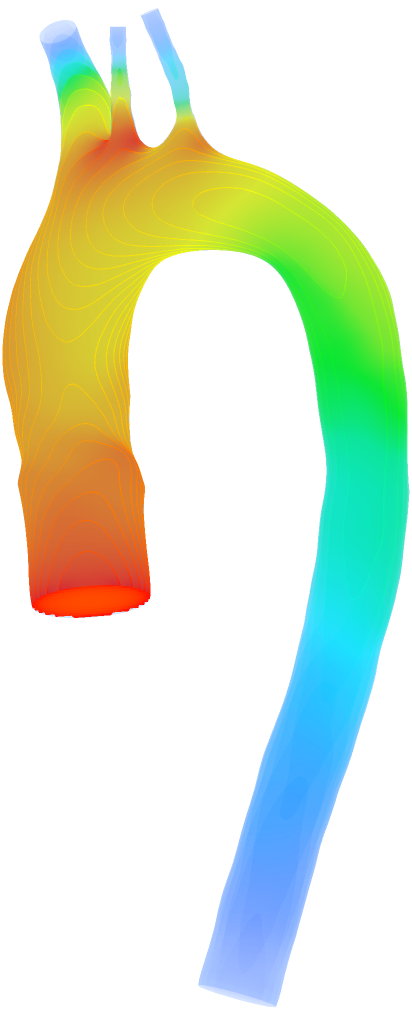}
        \caption{Diabetic}
        \label{pr_b}
    \end{subfigure}
    \hfill
    \begin{subfigure}{0.18\textwidth}
        \includegraphics[width=\linewidth]{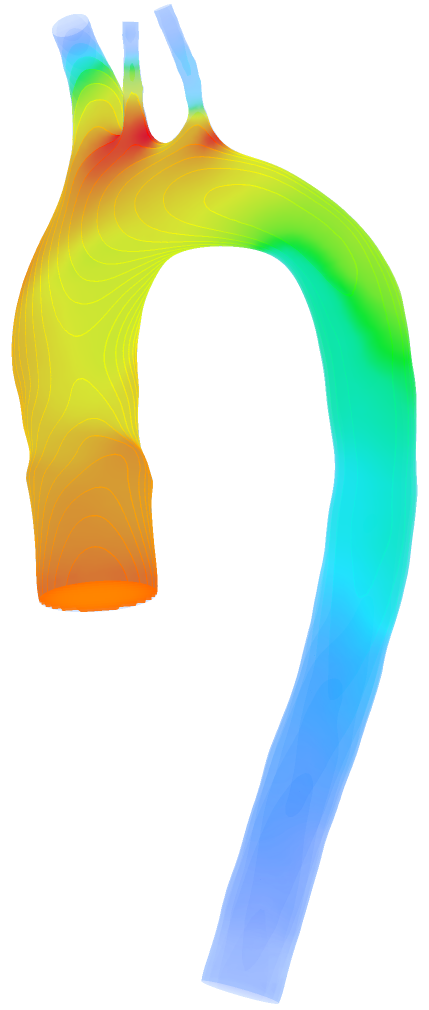}
        \caption{Healthy 1}
        \label{pr_c}
    \end{subfigure}
    \hfill
    \begin{subfigure}{0.18\textwidth}
        \includegraphics[width=\linewidth]{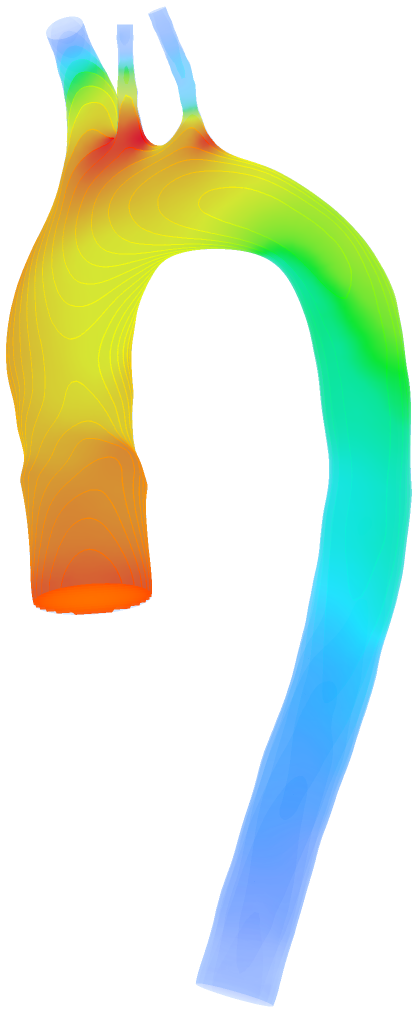}
        \caption{Healthy 2}
        \label{pr_d}
    \end{subfigure}
    \hfill
    \begin{subfigure}{0.08\textwidth}
        \includegraphics[width=\linewidth]{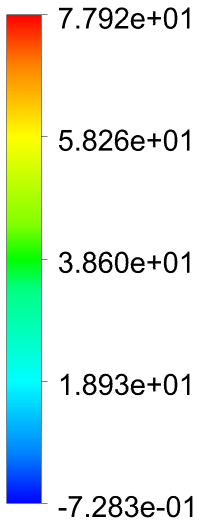}
        \caption*{}
    \end{subfigure}
    \captionsetup{justification=justified, singlelinecheck=false}
    \caption{Volume-averaged pressure contours in the aortic arch for different blood flow cases: (a) anemic, (b) diabetic, (c) healthy (Case 1), and (d) healthy (Case 2). Red regions correspond to areas of higher pressure, while blue regions indicate lower pressure, with values presented in Pascal (Pa).}
    \label{fig:pr}
\end{figure}

Figures~\ref{fig:pc} and~\ref{fig:pr} illustrate the pressure contours for different blood conditions, comparing surface pressure (Figure~\ref{fig:pc}) and volume-averaged pressure (Figure~\ref{fig:pr}) distributions across four cases: anemic, diabetic, healthy (Case 1), and healthy (Case 2). The pressure patterns highlight the influence of blood viscosity, hematocrit levels, and vascular geometry on the overall flow resistance in the aortic arch.

In the anemic case (Figures~\ref{pc_a} and~\ref{pr_a}), the lower viscosity and hematocrit result in smoother flow with reduced pressure gradients across the aortic arch. The ascending aorta shows relatively uniform pressure, with minimal pressure drops near the outlets (brachiocephalic trunk, left carotid artery, and left subclavian artery). This reflects reduced resistance to flow and lower energy dissipation due to the lower viscosity of anemic blood. The descending aorta also displays smooth pressure variations, emphasizing the reduced overall flow resistance associated with this condition. Clinically, such low pressures may be associated with reduced perfusion, especially in microcirculation, and may predispose patients to orthostatic intolerance or inadequate capillary oxygen delivery~\cite{cines1998endothelial}.

In contrast, the diabetic case (Figures~\ref{pc_b} and~\ref{pr_b}), in contrast, demonstrates significantly elevated pressure near the ascending aorta and along the inner curvature of the aortic arch. The higher viscosity and hematocrit of diabetic blood lead to amplified velocity gradients, increasing flow resistance and resulting in pronounced pressure drops near the branch outlets. The brachiocephalic trunk and left carotid artery exhibit the most substantial pressure differences, reflecting the greater energy loss associated with the more viscous fluid. The descending aorta also experiences higher pressure compared to the anemic case, highlighting the systemic effects of increased flow resistance in diabetic blood. These features are consistent with clinical observations of increased afterload and impaired pressure-buffering in diabetic vasculature~\cite{Sun2014}.

Healthy cases (Figures~\ref{pc_c}, \ref{pc_d}, \ref{pr_c}, and~\ref{pr_d}) fall between the extremes of anemic and diabetic conditions. In Case 1, the pressure distribution near the ascending aorta and branch outlets is slightly higher than in Case 2, reflecting the moderate influence of hematocrit and viscosity. Both cases show localized high-pressure regions at the branch openings, indicative of flow acceleration and geometric effects, but these are more subdued compared to the diabetic case. The descending aorta in both healthy cases displays relatively smooth pressure variations, consistent with balanced hemodynamic conditions.

The comparison across cases underscores the role of blood rheology in shaping pressure distributions and flow resistance in the aortic arch. Anemic blood, with its lower viscosity, minimizes pressure drops and energy dissipation, leading to smoother flow patterns. Diabetic blood, on the other hand, imposes greater resistance, with higher pressure gradients and localized energy losses near bifurcations and outlets. Healthy cases strike a balance between these extremes, reflecting physiological conditions that maintain efficient blood flow while accommodating the natural geometric complexity of the aortic arch.

\begin{figure}[h!]
    \centering
    \begin{subfigure}{0.45\textwidth}
        \includegraphics[width=\linewidth]{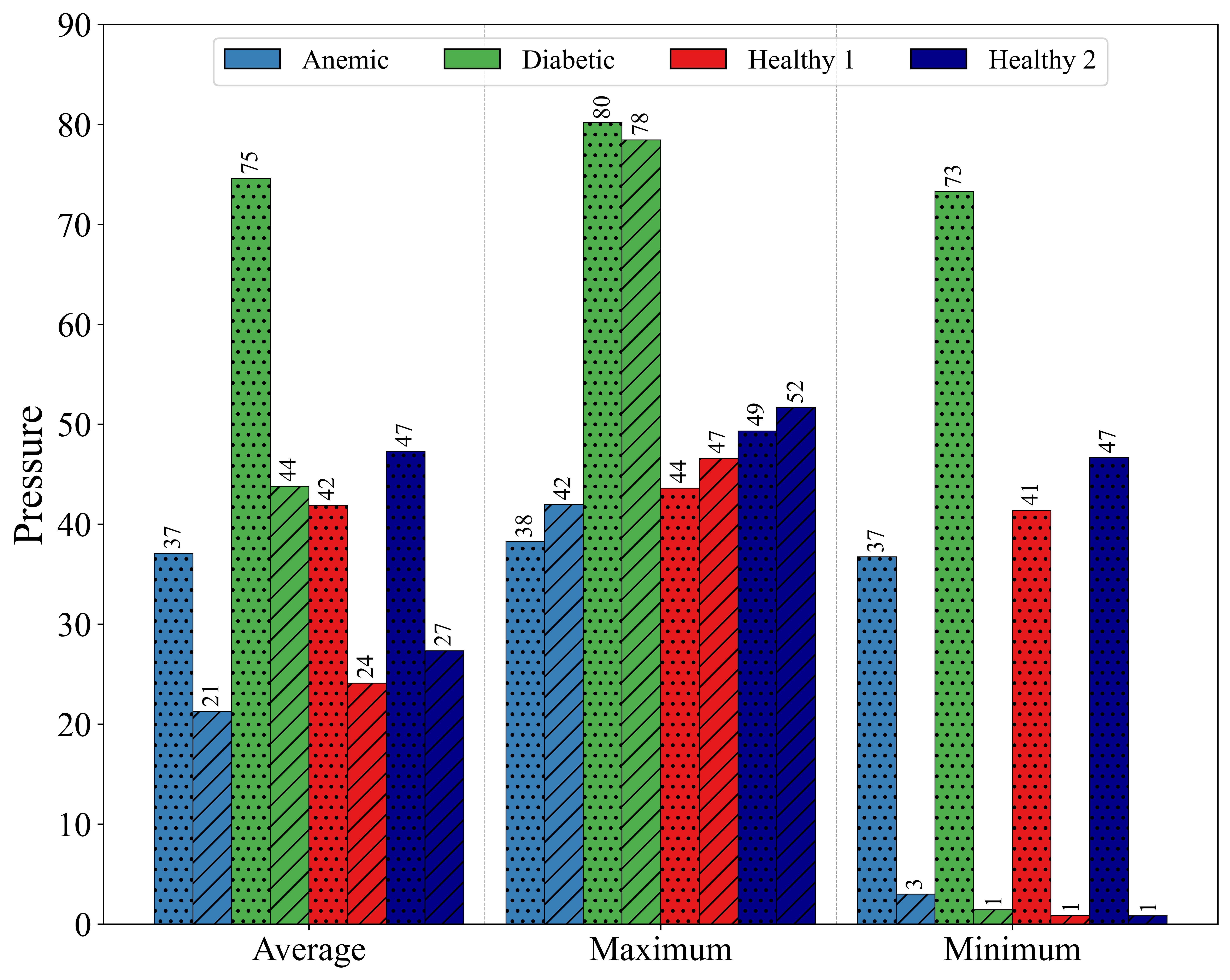}
        \caption{Pressure distribution}
        \label{fig:pressure-distribution}
    \end{subfigure}
    \hfill
    \begin{subfigure}{0.45\textwidth}
        \includegraphics[width=\linewidth]{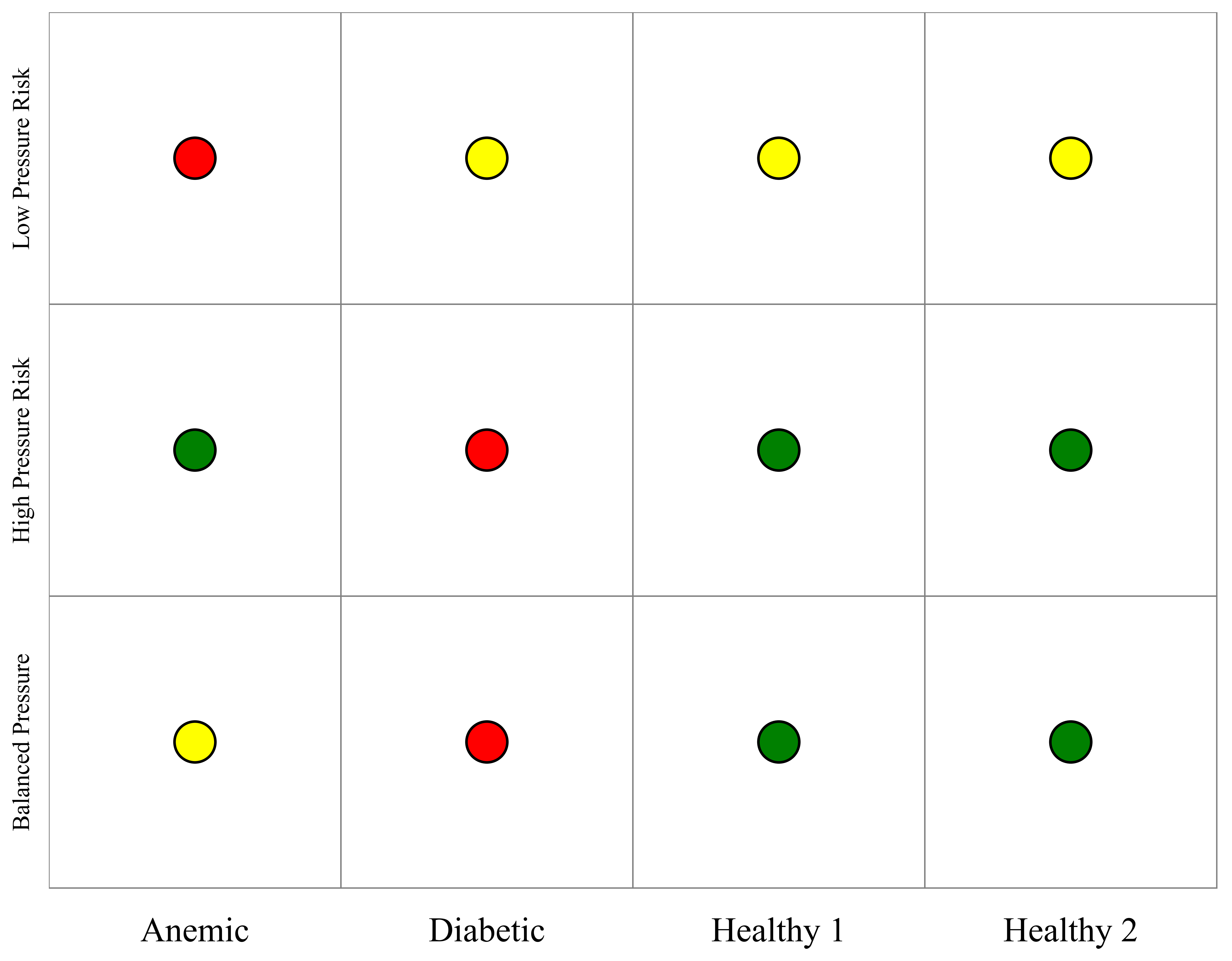}
        \caption{Pressure-based risk classification}
        \label{fig:pressure-risk}
    \end{subfigure}
    \caption{(a) Pressure distribution across four blood conditions—anemic, diabetic, healthy (Case 1), and healthy (Case 2)—showing average, maximum, and minimum values. \textcolor{blue}{Dotted bars} indicate inlet pressure, while \textcolor{blue}{diagonally hatched bars} represent wall pressure. Diabetic cases exhibit elevated pressure magnitudes at both inlet and wall locations, while anemic cases show considerably lower wall pressure values. (b) Pressure-based risk classification using three metrics: minimum, maximum, and average pressure. Color-coded circles represent risk levels, where \textcolor{blue}{red} indicates high risk, \textcolor{blue}{yellow} indicates borderline or moderate concern, and \textcolor{blue}{green} indicates no risk.}
\end{figure}
Figure~\ref{fig:pressure-distribution} presents a quantitative comparison of inlet and wall pressures for all cases, based on maximum, average, and minimum values. Diabetic blood exhibits the highest pressures across all metrics, particularly at the wall, reflecting increased flow resistance due to elevated viscosity. In contrast, the anemic case consistently shows the lowest pressure values, which aligns with reduced hematocrit and energy dissipation. Healthy Cases 1 and 2 fall between these extremes, with Case 1 showing slightly higher pressures, possibly due to modest rheological differences. These pressure patterns corroborate the flow observations and provide a basis for evaluating systemic load and perfusion potential.

Figure~\ref{fig:pressure-risk} provides a categorical risk assessment based on minimum, maximum, and average wall pressure metrics. The classification logic is physiologically informed and intended to capture deviations from optimal hemodynamic conditions.\footnote{In pulsatile aortic blood flow, the minimum and maximum luminal pressures could be interpreted as approximations of diastolic and systolic pressures, respectively. Although CFD simulations typically do not model arterial wall compliance unless coupled with structural mechanics, these pressure extrema still offer useful proxies for clinical pressure phases when performing comparative risk analysis.} The classification matrix emphasizes key clinical contrasts. The diabetic profile is flagged for high-pressure risk due to its elevated maximum wall pressure, consistent with increased vascular resistance and strain on the arterial wall. It also shows a borderline response in minimum and average pressure, suggesting incomplete normalization even during diastolic phases~\cite{Safar2017}. The anemic condition is marked by low-pressure risk, particularly in the minimum metric, which reflects reduced circulatory force and potential for inadequate capillary perfusion. Both healthy cases—particularly Healthy 2—fall predominantly in the no-risk zone, reinforcing the interpretation that these conditions exhibit stable, well-regulated hemodynamics supportive of systemic vascular health.

\subsection{Wall Shear Stress Distribution and Risk Stratification}

Wall shear stress (WSS) is a critical hemodynamic quantity representing the tangential force exerted by blood flow on vascular endothelium. It plays a central role in regulating endothelial function, vascular remodeling, and the development of atherosclerosis. Clinical and experimental studies have shown that deviations from physiological WSS values are associated with endothelial dysfunction, plaque formation, and thrombotic risk~\cite{Malek1999, Chiu2011, Giddens1993}. In particular, low WSS promotes leukocyte adhesion, platelet aggregation, and pro-inflammatory signaling, while abnormally high WSS may induce endothelial denudation and arterial remodeling~\cite{Stone2012}.
\begin{figure}[h!]
    \centering
    \begin{subfigure}{0.18\textwidth}
        \includegraphics[width=\linewidth]{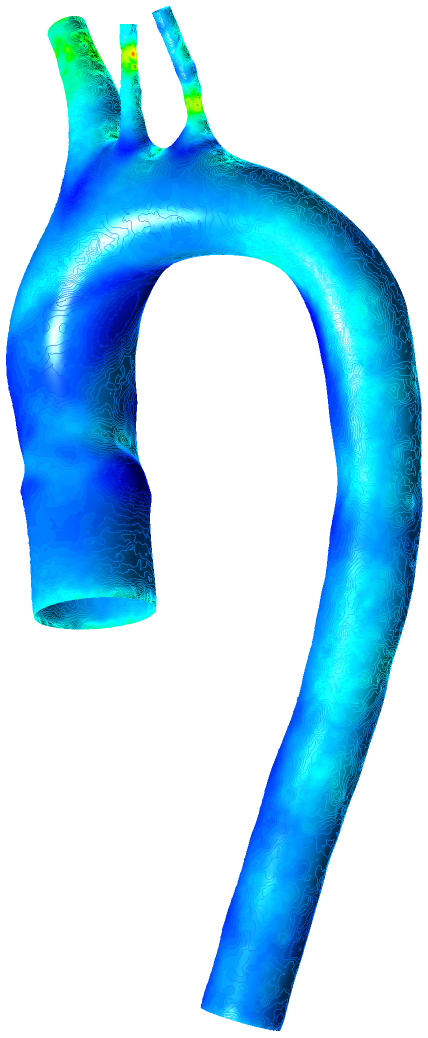}
        \caption{Anemic}
        \label{wss_a}
    \end{subfigure}
    \hfill
    \begin{subfigure}{0.18\textwidth}
        \includegraphics[width=\linewidth]{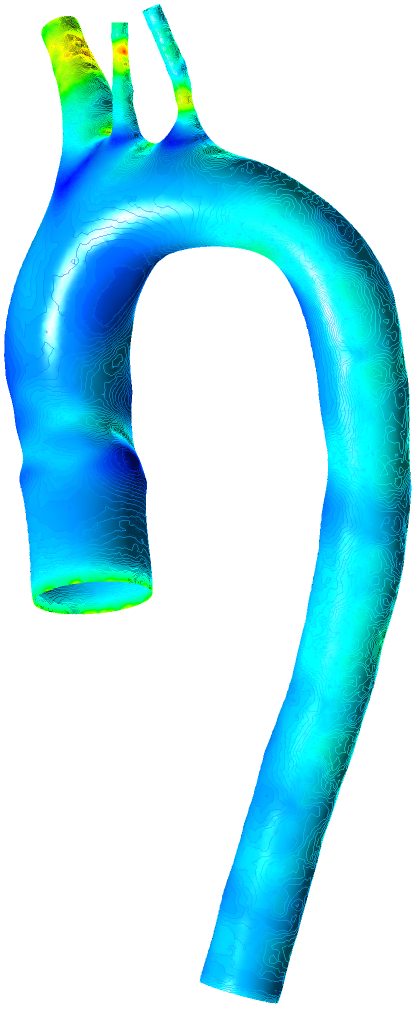}
        \caption{Diabetic}
        \label{wss_b}
    \end{subfigure}
    \hfill
    \begin{subfigure}{0.18\textwidth}
        \includegraphics[width=\linewidth]{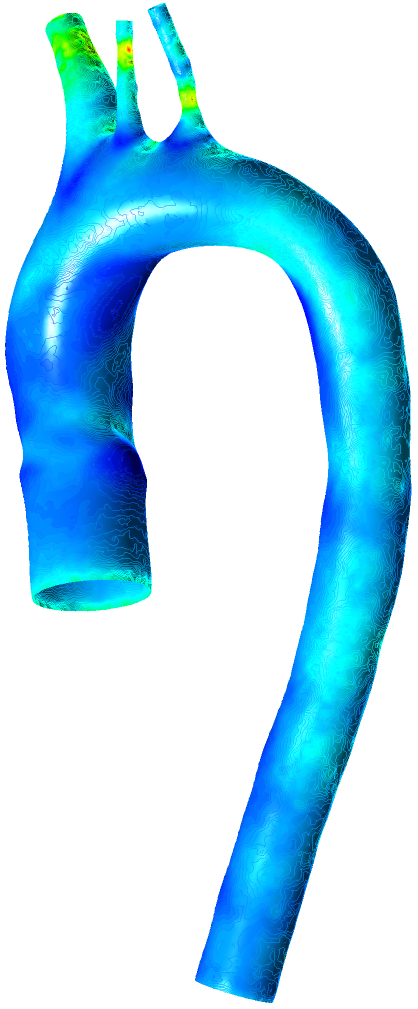}
        \caption{Healthy 1}
        \label{wss_c}
    \end{subfigure}
    \hfill
    \begin{subfigure}{0.18\textwidth}
        \includegraphics[width=\linewidth]{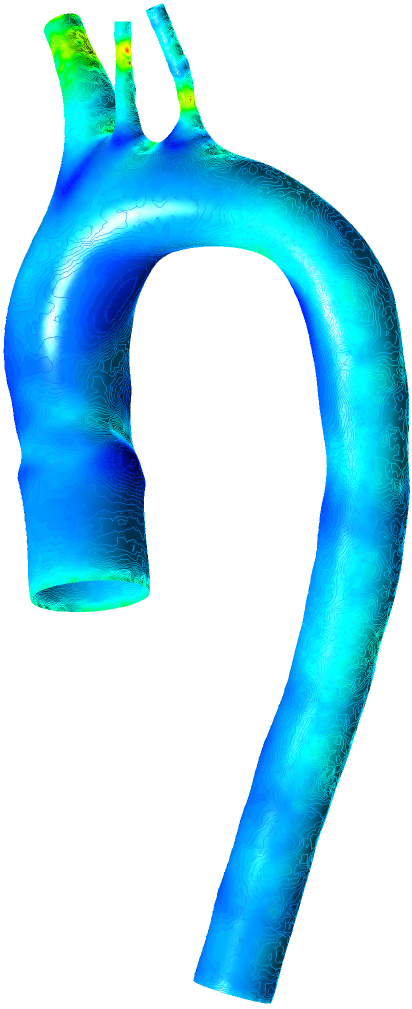}
        \caption{Healthy 2}
        \label{wss_d}
    \end{subfigure}
    \hfill
    \begin{subfigure}{0.08\textwidth}
        \includegraphics[width=\linewidth]{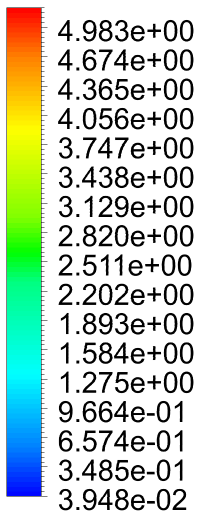}
    \end{subfigure}
    \caption{Wall shear stress (WSS) contours in the aortic arch for different blood flow cases: (a) anemic, (b) diabetic, (c) healthy (Case 1), and (d) healthy (Case 2). Red regions correspond to higher WSS values, while blue regions indicate lower WSS values in Pascal (Pa).}
    \label{fig:wss}
\end{figure}

Figure~\ref{fig:wss} shows the spatial distribution of WSS under four blood conditions. The anemic case exhibits smooth, uniform flow with moderate WSS magnitudes throughout the aortic arch. This is attributed to low hematocrit and viscosity, which reduce flow resistance and minimize sharp velocity gradients. While such flow may reduce mechanical stress on the endothelium, persistently low WSS may impair shear-mediated signaling pathways, such as nitric oxide synthesis, leading to altered endothelial function~\cite{Chakraborty2013}.

In contrast, the diabetic case shows elevated and highly non-uniform WSS, especially near the brachiocephalic trunk and left carotid artery. This arises from increased viscosity and reduced arterial compliance, both of which intensify velocity gradients in complex flow regions. These observations align with clinical reports of endothelial stress and vascular remodeling in diabetes~\cite{Steinman2003}. The healthy cases lie between these two extremes: both show moderate WSS values with localized elevations near outlets due to physiological flow acceleration and geometric curvature.

\begin{figure}[h!]
    \centering
    \begin{subfigure}{0.45\textwidth}
        \includegraphics[width=\linewidth]{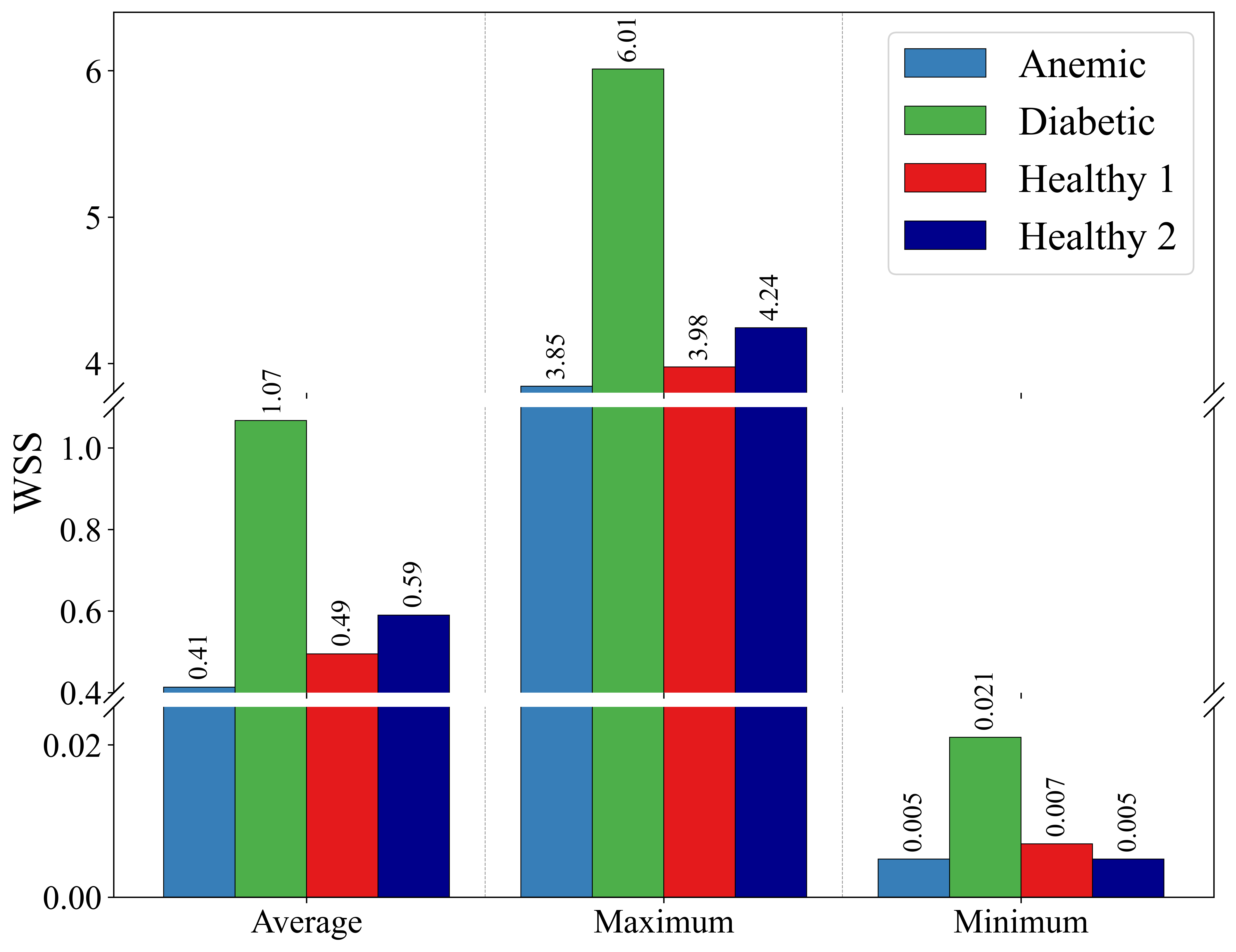}
        \caption{WSS distribution}
        \label{fig:wss-metrics}
    \end{subfigure}
    \hfill
    \begin{subfigure}{0.45\textwidth}
        \includegraphics[width=\linewidth]{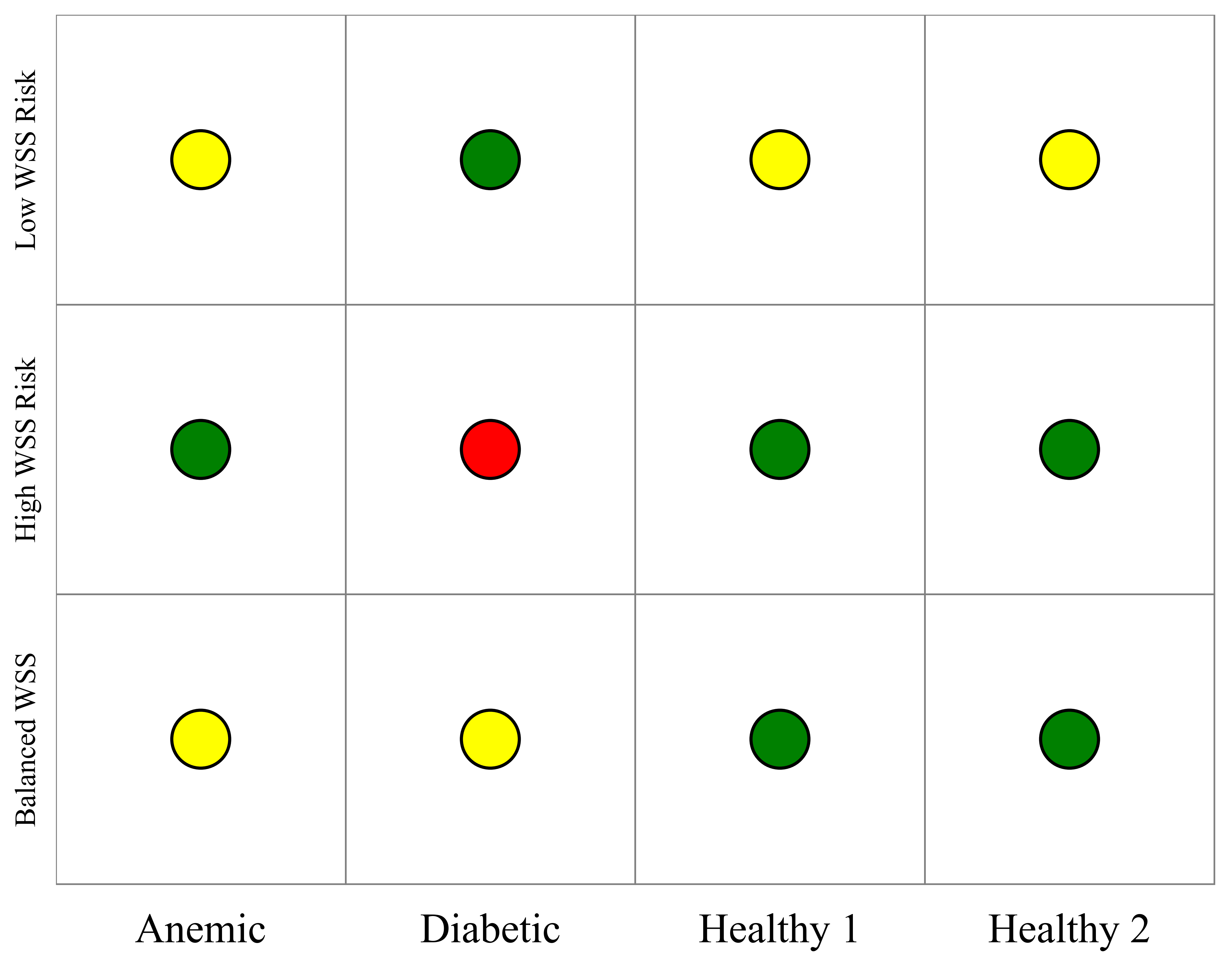}
        \caption{WSS-based risk classification}
        \label{fig:wss-risk}
    \end{subfigure}
    \caption{(a) WSS distribution across four blood conditions--anemic, diabetic, healthy (Case 1), and healthy (Case 2)--showing average, maximum, and minimum values. A broken y-axis is used to accommodate both low (below 0.05~Pa) and high (above 1~Pa) WSS values. (b) Risk classification based on three WSS metrics: minimum, maximum, and average WSS. Each cell is color-coded to represent the severity of the condition under a given metric, where \textcolor{blue}{red} indicates high risk, \textcolor{blue}{yellow} indicates borderline or moderate concern, and \textcolor{blue}{green} indicates no risk.}
    \label{fig:wss-comparison}
\end{figure}

Figure~\ref{fig:wss-comparison}(a) quantifies these observations. The diabetic case yields the highest maximum WSS (6.01~Pa), exceeding physiological thresholds, while the anemic case shows the lowest average (0.041~Pa) and minimum (0.005~Pa) WSS. Healthy cases remain mostly within the accepted physiological range of 0.4 to 1.5~Pa~\cite{Malek1999}. Panel (b) translates these metrics into a color-coded risk map: the diabetic case shows high-WSS risk (red), while the anemic case approaches the low-WSS threshold (yellow). The healthy cases are within the safe zone (green) across all three metrics.

These results confirm that blood rheology modulates the local hemodynamic environment, with clinical implications for cardiovascular risk. Anemia leads to sub-threshold WSS that may impair mechanosensitive endothelial responses, whereas diabetes imposes excessive WSS variations that can cause mechanical endothelial injury. These findings reinforce the role of WSS as a predictive biomarker for vascular dysfunction and support its clinical application in disease risk stratification~\cite{Sun2014, Stone2012}. From a translational perspective, identifying regions of abnormal WSS can inform patient-specific risk for intimal thickening, aneurysmal remodeling, or thrombus initiation, particularly in high-risk groups such as diabetics and anemic individuals~\cite{Akinsheye2010, Diamond2016}.

Overall, the comparative analysis across four representative blood conditions highlights the critical role of rheological variation in modulating aortic hemodynamics. Velocity fields, pressure distributions, and wall shear stress patterns each reveal distinct mechanistic signatures associated with anemic, diabetic, and healthy flows. These signatures not only reflect energy transport and flow efficiency but also indicate biomechanical environments that may promote endothelial dysfunction, thrombus formation, or vascular remodeling. The integration of simulation-derived hemodynamic metrics with established clinical risk factors could provide a non-invasive framework for stratifying cardiovascular risk and guiding personalized intervention strategies~\cite{Diamond2016, Safar2017}.
\FloatBarrier

\section{Conclusion} \label{sec:conclusion}

This study presented a comprehensive computational framework to examine blood flow dynamics in the aortic arch across four physiological and pathological cases: anemic, diabetic, and two healthy conditions. By employing the Carreau model for non-Newtonian viscosity, the simulations captured realistic blood rheology and enabled detailed analysis of velocity profiles, pressure fields, and wall shear stress distributions. The results underscore the importance of integrating blood properties, vessel geometry, and boundary conditions to understand vascular mechanics and their clinical consequences.

Key findings show that anemic flow, characterized by low hematocrit and viscosity, resulted in smooth velocity streamlines, reduced pressure gradients, and uniformly low WSS values. Although energetically efficient, this condition may limit endothelial stimulation and impair perfusion, particularly in downstream or microvascular territories. Diabetic blood flow, in contrast, exhibited elevated viscosity and shear, which led to steeper velocity gradients, heightened WSS magnitudes, and significant pressure drops at outlet branches. These mechanical features are closely associated with pro-atherogenic stimuli, endothelial dysfunction, and increased cardiac workload, thereby validating the clinical association between abnormal blood rheology and vascular risk.

The two healthy cases demonstrated intermediate flow behavior, with velocity and WSS remaining largely within physiologic ranges. Flow acceleration near bifurcations and mild pressure asymmetries reflected normal anatomical constraints rather than pathology. Importantly, the study showed that hemodynamic indicators such as localized WSS peaks or recirculation zones can serve as proxies for disease-prone regions—even in geometries not yet anatomically compromised. These patterns reveal the systemic consequences of altered blood properties, offering mechanistic insight into how seemingly mild changes in rheology may initiate or amplify cardiovascular stress across different regions of the aorta.

Despite these insights, the study has certain limitations. The simulations were conducted on a generalized aortic geometry rather than patient-specific models, which may limit anatomical specificity. Blood flow was assumed to be laminar and incompressible, with rigid arterial walls—thus excluding potential contributions from fluid-structure interaction and wall elasticity. Additionally, while the Carreau model captures shear-thinning effects, it does not account for pathological viscoplasticity seen in conditions like thrombosis or inflammation. Experimental validation was beyond the scope of this work but remains essential to confirm key predictions, particularly regarding shear stress thresholds and risk classification.

Future research should address these limitations by integrating patient-derived geometries, incorporating compliant wall models through FSI, and expanding simulations to pathological cases such as stenosis and aneurysm. Incorporating alternative rheological models like Herschel–Bulkley or Casson could improve accuracy in diseased states. Coupling CFD with thrombotic and inflammatory pathways will enable richer physiological insights, while machine learning frameworks may enhance speed and predictive capability. Ultimately, this modeling approach can contribute to non-invasive vascular risk stratification and support the design of personalized therapeutic strategies.


\section*{Declaration of Interests}
The authors declare that they have no known competing financial interests or personal relationships that could have appeared to influence the work reported in this paper.

\section*{Author contributions}
F.A. Tina: conceptualization, data curation, formal analysis, investigation, methodology, resources, software, validation, visualization, writing -- original draft, writing -- review \& editing. H. Ahmed: conceptualization, formal analysis, investigation, methodology, validation, visualization, supervision, writing -- original draft, writing -- review \& editing.  H. R. Biswas:  supervision, writing -- review \& editing.

\section*{Data Availability Statement}
The data that support the findings of this study are available upon request to the corresponding author.

\section*{Funding}  
Self-funded.


\end{document}